\documentstyle[epsf,epsfig,mprocl]{article}

\begin{document}

\title{TESTING A NEW MONTE CARLO STRATEGY FOR FOLDING MODEL PROTEINS } 

\author{HELGE FRAUENKRON$^1$, UGO BASTOLLA$^1$, ERWIN GERSTNER$^{1, 2}$,\\ 
      PETER GRASSBERGER$^{1, 2}$, AND WALTER NADLER$^1$}
\address{ $^1$ HLRZ c/o Forschungszentrum J\"ulich,\\ D-52425 J\"ulich,
      Germany\\
$^2$ Physics Department, University of Wuppertal,\\ D-42097 Wuppertal, Germany}

\date{\today}

\maketitle

\abstracts{
  We demonstrate that the recently proposed pruned-enriched
  Rosenbluth method PERM (P.~Grassberger, Phys.~Rev.~{\bf E 56} (1997)
  3682) leads to very efficient algorithms for the folding of simple model
  proteins. We test it on several models for lattice heteropolymers, and
  compare to published Monte Carlo studies of the properties of particular
  sequences. In {\it all} cases our method is faster than the previous
  ones, and in several cases we find new minimal energy states. In
  addition to producing more reliable candidates for ground states, our
  method gives detailed information about the thermal spectrum and, thus,
  allows to analyze static aspects of the folding behavior of arbitrary
  sequences.  
}

\section{Introduction}
\label{sec:intro}

Protein folding
\cite{Gierasch:King:90,Creighton:92,Merz:LeGrand:94,Brunak:95}
is one of
the most interesting and challenging problems in polymer physics and
mathematical biology.  It is concerned with the problem of how a
heteropolymer of a given sequence of amino acids folds into precisely that
geometrical shape in which it performs its biological function as a
molecular machine.\cite{Drexler:86,Gross:95} Currently, it is much
simpler to find coding DNA---and, thus, also amino acid---sequences
than to elucidate the 3-$d$ structures of given proteins.  Therefore,
solving the protein folding problem would be a major break-through in
understanding the biochemistry of the cell and, furthermore, in designing
artificial proteins.

In this contribution we are concerned with the direct approach: given a
sequence of amino acids, a molecular potential, and no other information,
find the ground state and the equilibrium state at physiological
temperatures.  Note that we are not concerned with the kinetics of folding,
but only in the final outcome. Also, we will not address the problems of
how to find good molecular
potentials,\cite{Crippen:Maiorov:94,Kolinski:Skolnick:95,Mirny:Shakhnovich:96}
and what is the proper level of detail in describing
proteins.\cite{Kolinski:Skolnick:95} Instead, we will use simple
coarse-grained models which have been proposed in the literature and have
become standards in testing the efficiency of folding algorithms.

Lots of methods have been proposed to solve this problem, ranging from
simple Metropolis Monte Carlo simulations at some nonzero temperature
\cite{Sali:Shakhnovich:Karlplus:94:JMB} over multi-canonical simulation
approaches \cite{Hansmann:Okamoto:94-96} to stochastic optimization schemes
based, e.g., on simulated annealing,\cite{Wilson:Cui:94} and genetic
algorithms.\cite{Unger:Moult:93,LeGrand:Merz:94a} Alternative methods use
heuristic principles,\cite{Dill:Fiebig:Chan:93} information from databases
of known protein structures,\cite{Eisenhaber:Persson:Argos:95} sometimes
in combination with known physico-chemical properties of small peptides.

The algorithms we apply here are variants of the pruned-enriched Rosenbluth
method (PERM).\cite{alg} This is a chain growth approach based on the
Rosenbluth-Rosenbluth (RR) \cite{rr} method. We will provide details on the
algorithm and on the analyses that can be performed, and we will present
more detailed results on ground state and spectral properties, and on the
folding behavior of the sequences analyzed.

\section{The Models}
\label{sec:models}

The models we study in this contribution are heteropolymers that live on 2-
and 3-dimensional regular lattices. They are self-avoiding chains with
attractive or repulsive interactions between neighboring non-bonded
monomers.

The majority of authors considered only two kinds of monomers. Although
also different interpretations are possible for such a binary choice, e.g.
in terms of positive and negative electric charges,\cite{kantor-kardar:94}
the most important model of this class is the HP
model.\cite{Dill:85,Dill:89-91} There, the two monomer types are assumed to
be hydrophobic (H) and polar (P), with energies $\epsilon_{HH}=-1,\;
\epsilon_{HP}=\epsilon_{PP}=0$ for interaction between not covalently bound
neighbors. Since this parameter set leads to highly degenerate ground
states, alternative parameters were proposed, e.g.
$\boldmath{\epsilon}=(-3,-1,-3)$ \cite{socci1} and
$\boldmath{\epsilon}=(-1,0,-1)$.\cite{otoole} Note, however, that in these
latter parameter sets, since they are symmetric upon exchange of H and P,
the intuitive distinction between hydrophilic and polar monomers gets lost.

In the other extremal case of models, all monomers of a sequence are
considered to be different, and interaction energies are drawn randomly
from a continuous
distribution.\cite{Sali:Shakhnovich:Karplus:94:Nature,Klimov:Thirumalai:96}
These models correspond, effectively, to assuming an infinite number of
monomer types.

\section{The Algorithm}
\label{sec:algorithm}

The algorithms we apply here are variants of the pruned-enriched Rosenbluth
method (PERM),\cite{alg} a chain growth algorithm based on the
Rosenbluth-Rosenbluth (RR) \cite{rr} method. Monomers are added
sequentially, the $n$-th monomer being placed at site $i$ with probability
$p_n(i)$. In {\it simple sampling}, $p_n(i)$ is uniform on all neighbors of
the last monomer, leading to exponential attrition. The original RR method
avoids this by using a uniform $p_n(i)$ on all {\it vacant} neighbors of
$i_{n-1}$. More generally, we call any non-uniform choice of $p_n(i)$ a
generalized RR method.  The relative thermal weight of a particular chain
conformation of length $n$ is then determined by $ W_n = m_n
\exp(-\beta\Delta E_n) W_{n-1}$, with $W_1=1$; $\Delta E_n$ is the energy
gain from adding monomer $n$; and $m_n$ is the Rosenbluth factor,
$m_n=\sum_{j \in \left\{\rm nn\right\} } p_n(j)/p_n(i)$.  We note that
$W_n$ is also an estimate for the partition function $Z_n$ of the
$n$-monomer chain.\cite{alg} Chain growth is stopped when the final size
is reached and started new from $n=1$.

In easy cases, $p_n(i)$ can be chosen so that Boltzmann and Rosenbluth
factors---or Rosenbluth factors for different $n$---cancel, leading to
narrow weight distributions.  But in general, this algorithm produces a
wide spread in weights that can lead to serious problems.\cite{kremer} On
the other hand, since the weights accumulate as the chain grows, one can
interfere during the growth process by `pruning' conformations with low
weights and enriching high-weight conformations. This is, in principle,
similar to population based methods in polymer simulations
\cite{garel,velicson} and in quantum Monte Carlo (MC).\cite{umrigar}
However, our implementation is different. Pruning is done stochastically:
if the weight of a conformation has decreased below a threshold $W_n^<$, it
is eliminated with probability 1/2, while it is kept and its weight is
doubled in the other half of cases. Enrichment \cite{enrich} is done
independently of this: if $W_n$ increases above another threshold $W_n^>$,
the conformation is replaced by $n_c$ copies, each with weight $W_n/n_c$.
Technically, this is done by putting onto a stack all information about
conformations which still have to be copied. This is most easily
implemented by recursive function calls.\cite{alg} Thereby the need for
keeping large populations of conformations \cite{garel,velicson,umrigar} is
avoided. PERM has proven extremely efficient for studies of lattice
homopolymers near the $\theta$ point,\cite{alg} their phase
equilibria,\cite{multic} and of the ordering transition in semi-stiff
polymers.\cite{stiff}

The main freedom when applying PERM consists in the a priori choice of the
sites where to place the next monomer, in the thresholds $W^<$ and $W^>$
for pruning and copying, and in the number of copies made each time. All
these features do not affect the formal correctness of the algorithm, but
they can greatly influence its efficiency. They may depend arbitrarily on
chain lengths and on local configurations, and they can be changed freely
at any time during the simulation. Thus the algorithm can `learn' during
the simulation.

In order to apply PERM to heteropolymers at very low temperatures, 
the strategies proposed in Ref.\cite{alg} are modified as follows.

(1) For homopolymers near the theta-point it had been found that the best
choice for the placement of monomers was not according to their Boltzmann
weights, but uniformly on all allowed sites.\cite{alg,multic} This might
be surprising since the Boltzmann factor has then to be included into the
weight of the configuration, which might lead to large fluctuations.
Obviously, this effect is counterbalanced by the fact that larger Boltzmann
factors correspond to higher densities and thus to smaller Rosenbluth
factors.\cite{kremer}

For a heteropolymer this has to be modified, as there is no longer a unique
relationship between density and Boltzmann factor. In a strategy of
`anticipated importance sampling' we should preferentially place monomers
on sites with mostly attractive neighbors. Assume that we have two kinds of
monomers, and we want to place a type-$A$ monomer. If an allowed site has
$m_B$ neighbors of type $B$ ($B=H,P$), we select this site with a
probability $\propto 1+a_{AH}m_H+a_{AP}m_P$. Here, $a_{AB}$ are constants
with $a_{AB}>0$ for $\epsilon_{AB}<0$ and vice versa.

(2) Most naturally, $W^>$ and $W^<$ are chosen proportional to the
estimated partition sum $Z_n$.\cite{alg} This becomes inefficient at very
low $T$ since $Z_n$ will be underestimated as long as no low-energy state
is found. When this finally happens, $W^>$ is too small. Thus too many
copies are made which are all correlated but cost much CPU time.

This problem can be avoided by increasing $W^>$ and $W^<$ during
particularly successful `tours' \cite{alg} (a tour is the set of configurations
derived from a single start). But then also the average number of long
chains is decreased in comparison with short ones.  To reduce this effect
and to create a bias towards a sample which is flat in chain length, we
multiply by some power of $M_n/M_1$, where $M_n$ is the number of generated
chains of length $n$. With ${\cal N}(n)$ denoting the number of chains
generated during the current tour we used therefore
$$
   W^< = C\;Z_n\; [(1+{\cal N}(n)/M)(M_n+M)/(M_1+M)]^2 , 
$$
and $W^> = rW^<$. Here, $C$ is a constant of order unity, $r\approx 10$, 
and $M$ is a constant of order $10^4 - 10^5$.

(3) Creating only one additional copy at each enrichment event (as done in
Ref.\cite{alg}), cannot prevent the weights from exploding at very low $T$.
Thus we have to make several copies if the weight is large and surpasses
$W^>$ substantially. A good choice for the number of new copies created
when $W>W^>$ is int$[1+\sqrt{W/W^>}]$.

(4) Two special tricks were employed for `compact' configurations of the
2-$d$ HP model filling a square. First of all, since we know in this case
where the boundary should be, we added a bias for polar monomers to
actually be on that boundary, by adding an additional energy of $-1$ per
boundary site.  Note that this bias has to be corrected in the weights,
thus the final distributions are unaffected by it and unbiased. Secondly,
in two dimensions we can immediately delete chains which cut the free
domain into two disjoint parts, since they never can grow to full length.
In the present simulations, we checked for this by looking ahead one time
step. In spite of the additional work this was very efficient, since it
reduced considerably the time spent on dead-end configurations.

(5) In some cases we did not start to grow the chain from one end but from
a point in the middle. We grew first one half, and then the other.  Results
were averaged over all possible starting points.  The idea behind this is
that real proteins have folding nuclei,\cite{Matheson:Scheraga:78} and it
should be most efficient to start from such a nucleus. In some cases this
trick was very successful and speeded up the ground state search
substantially, in others not.  We take this observation as an indication
that in various sequences the end groups already provide effective
nucleation sites. This is e.g.  the case for the 80-mer with modified HP
interactions of Ref.\cite{otoole}. We also tried to grow the chain on both
sides simultaneously. However it turned out that this is not effective
computationally.\cite{Grassberger:unpublished}

(6) For an effective sampling of low-lying states the choice of simulation
temperature $T$ appears to be of importance. If it is too large, low-lying
states will have a low statistical weight and will not be sampled reliably.
On the other hand, if $T$ is too low, the algorithm becomes too greedy:
configurations which look good at first sight but lead to dead ends are
sampled too often, while low energy configurations, whose qualities become
apparent only at late stages of the chain assembly, are sampled rarely. Of
course they then get huge weights (since the algorithm is correct after
all), but statistical fluctuations become huge as well. This is in complete
analogy to the slow relaxation hampering more traditional (Metropolis type)
simulations at low $T$---note, however, that ``relaxation" in the proper
sense does not exist in the present algorithm.

In the cases we considered it turned out to be most effective to choose a
temperature that is below the collapse transition temperature (note,
however, that this transition is smeared out, see the results below) but
somewhat above the temperature corresponding to the structural transition
which leads to the native state.  This observation corresponds
qualitatively to the considerations of
Ref.\cite{Finkelstein:Gutin:Badretdinov:95}, although a quantitative
comparison appears not to be possible. In some cases it also helped to
reduce the `greediness' of the algorithm by not making any
pruning/enrichment during the first steps.

\section{The Sequences}
\label{sec:sequences}

For the above models various sequences were analyzed in the literature,
and in Ref.\cite{Frauenkron:etal:97,proteins:97}
we took these analyses as a test for our algorithm.
Here we will take a closer look at the properties of some
of the sequences that were considered there.

\begin{table}[b]
\vglue-5mm
\caption{Newly found lowest energy states for binary sequences with 
interactions $\epsilon = (\epsilon_{HH}, \epsilon_{HP}, \epsilon_{PP})$. 
Configurations are encoded as sequences of {\it r\/}(ight), {\it l\/}(eft),
{\it u\/}(p), {\it d\/}(own), {\it f\/}(orward), and {\it b\/}(ackward).}
\label{seq.struct.table}
\vglue2mm\footnotesize
\begin{center}
\begin{tabular}{|ccc|} \hline
  $N$ , $d$  &    sequence   & old ${E_{\rm min}}^{\rm Ref.}$ \\ 
  $\epsilon$ & configuration & new $E_{\rm min}$  \hfil       \\ \hline
 100 , 2 & 
$P_6HPH_2P_5H_3PH_5PH_2P_2(P_2H_2)_2PH_5PH_{10}-$&\\
 $(-1,0,0)$ & $PH_2PH_7P_{11}H_7P_2HPH_3P_6HPH_2$ & $-44$ \cite{pekney} \\
 & $r_6ur_2u_3rd_5luldl_2drd_2ru_2r_3(rulu)_2urdrd_2ru_3lur_3dld_2-$&\\&$rur_5d_3l_5uldl_2d_3ru_2r_3d_3l_2urul$ 
& $-47$ \\
 &
$rdldldrd_2r_2d_3l_2drdldr_2dl_2dl_2(urul)_2urur_2ul_2u_2l_2drd_2-$&\\&$lul_2uru_2r_2u_
4rul_3drd_3l_2d_2ldlu_2ru_2lu_3rd_2rdr$ & $-47$ \\
 &
$u_3r_3ur_5dl_4drd_2r_2ulur_4dr_2dld_2lu_2luld_2rdl_2ul_3dr_2dr_2d-$&\\&$rurdr_3d_2luld
_2lu_2l_2drd_2lulu_3l_2ul_2u_3ru$
& $-47$ \\
 &
$rdr_3dldl_3drdrur_2ur_2ulur_2urd_2(ldrd)_2l_2u_2ldl_2dr_2dr_3dl_2d-$&\\&$lulul_2ul_2d_
3ldr_3u_2rdrdldldr_2urur_4u_3rul$ & $-47$ \\
 &
$rd_3rdldl_2uld_3ld_2rur_3dl_2dr_6ul_3u_2l_3ur_3ur_2dld_2rur_2ulu_2-$&\\&$l_2u_3lu_3r_2
d_5r_2d_2rdru_2lu_2r_2u_2ldl_2dl$
& $-47$ \\
 &
$r_3u_3ru_2ru_3l_2ur_2ul_2urul_4dr_2dldrdld_2rur_2dld_2luld_2rdl_3-$&\\&$uru_2ldld_3l_3
ururu_2rur_2ululdldl_2u_3ld_4r$ & $-47$ \\\hline

 100 , 2 &
$P_3H_2P_2H_4P_2H_3(PH_2)_3H_2P_8H_6P_2H_6P_9HPH_2PH_{11}-$&\\
 $(-1,0,0)$ & $P_2H_3PH_2PHP_2HPH_3P_6H_3$ & $-46$ \cite{pekney} \\
 & $ru_2ldlu_2ld_2lu_2lurulur_2d_2ru_4r_2dld_4ru_3rdrdld_6rdru_3rul_2u_2-$&\\&$rdr_2ululu_2
(rd)_3rur_2dldld_4lulu_2ru$
& $-49$ \\
 &
$u_3rdru_2rd_2ru_2r_2u_2ldluldlu_5ld_6l_2d_2lu_3r_2u_6l_2d_3ldr_2dl_2-$&\\&$dldlu_2ru_2l
dl_2drdl_2d_2rurdrd_3ru_3ru$
& $-49$ \\
 &
$ul_2drdl_2u_3ld_4ldrdl_2u_2l_2d_3l_2uru_3r_2u_3rd_3ru_4rul_5dldr_2-$&\\&$d_2luldldrdldl
u_3lul_2ulur_2dr_2u_3rd_4l$
& $-49$ \\ \hline


  80 , 3 & 
 ${PH_2P_3(H_3P_2H_3P_3H_2P_3)_3H_4P_4(H_3P_2H_3P_3H_2P_3)H_2}$ & $-94$ \cite{otoole,deutsch} \\
 $(-1,0,-1)$ & 
$lbruflbl_2br_2drur_2dldl_3ulfrdr_3urfldl_3ulurur_3drblul_3-$&\\&$br_3bl_3dldrdr_3urul_2dlu 
$ & $-98$  \\ \hline
\end{tabular}
\end{center}
\end{table}

\subsection{2d HP model}
\label{sec:sequences:2dHP}

\begin{figure}[t]
  \begin{center}
    \psfig{figure=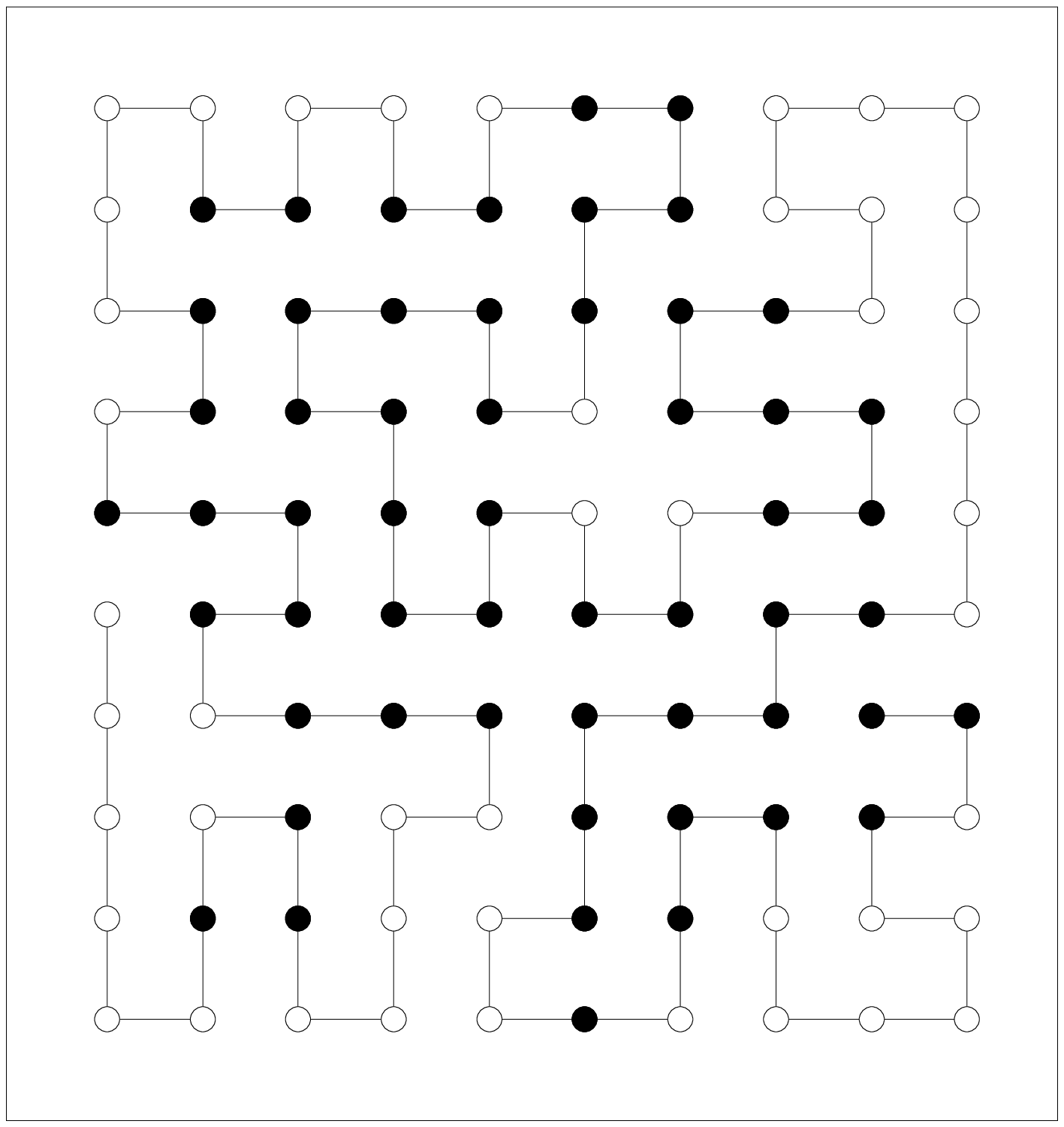, width=2.0truein}
    \hfil
    \psfig{figure=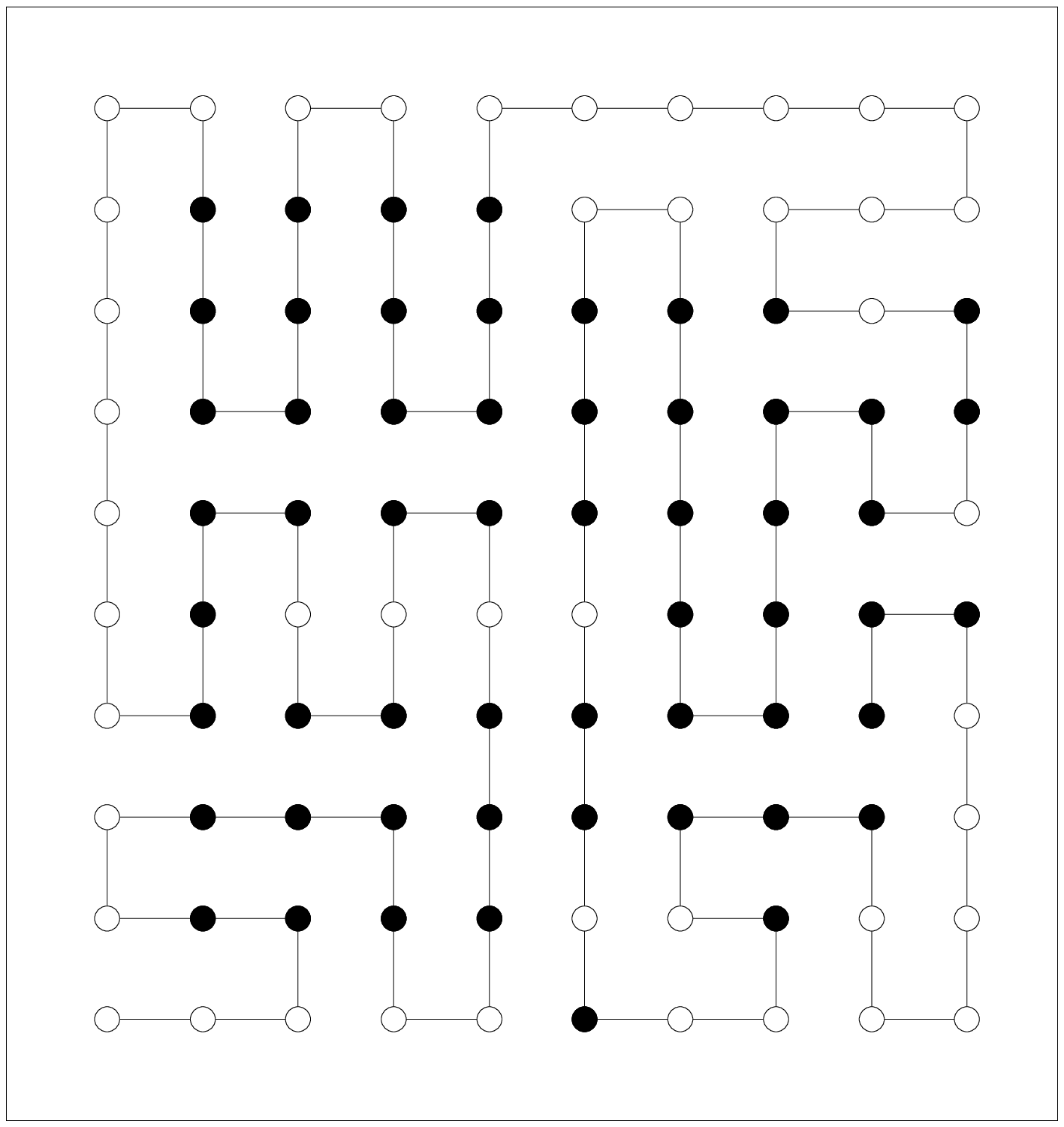, width=2.0truein}
  \end{center}
  \begin{center}
    \vglue-5mm
    \begin{minipage}[t]{6truecm}
\caption{ Putative compact native structure of sequence 1 from Table~1 ($E=-45$)
according to Ref.\protect\cite{pekney};
(filled circle) H monomers, (open circle) P monomers. \label{rama1.struct}}
    \end{minipage}
    \hfil
    \begin{minipage}[t]{6truecm}
      \caption{ Putative compact native structure of sequence 2 from Table~I
        ($E=-46$) according to Ref.\protect\cite{pekney}.\label{rama2.struct}}
    \end{minipage}
  \end{center}
  \vglue-5mm
\end{figure}

Two-dimensional HP chains were used in several papers as test cases for
folding algorithms. Two chains with $N=100$ were studied in
Ref.\cite{pekney}. The authors claimed that their native configurations
were compact, fitting exactly into a $10\times 10$ square, and had energies
$-44$ and $-46$, see Table~1 for the sequences and Fig.~\ref{rama1.struct}
and \ref{rama2.struct} for the respective proposed ground state structures.
These conformations were found by a specially designed MC algorithm which
should be particularly efficient for compact configurations.

\subsection{3d HP model}
\label{sec:sequences:3dHP}

Ten sequences of length $N=48$ were given in Ref.\cite{yue-shak}.  Each of
these sequences was designed by minimizing the energy of a particular
target conformation in sequence space under the constraint of constant
composition.\cite{Shakhnovich:93-94} The authors tried to find the lowest
energy states with two different methods, one being an heuristic stochastic
approach,\cite{Dill:Fiebig:Chan:93} the other based on exact enumeration
of low energy states.\cite{Yue:Dill:95}  With the first method they failed
in all but one case to reach the lowest energy. With the second method in
all but one cases they obtained conformations with energies that were even
lower than the putative ground states the sequences were designed for,
while for one case the ground state energy was confirmed.  Precise CPU
times were not quoted.

\subsection{3d modified HP model}
\label{sec:sequences:3dmodHP}

\begin{figure}
  \begin{center}
    \epsfig{figure=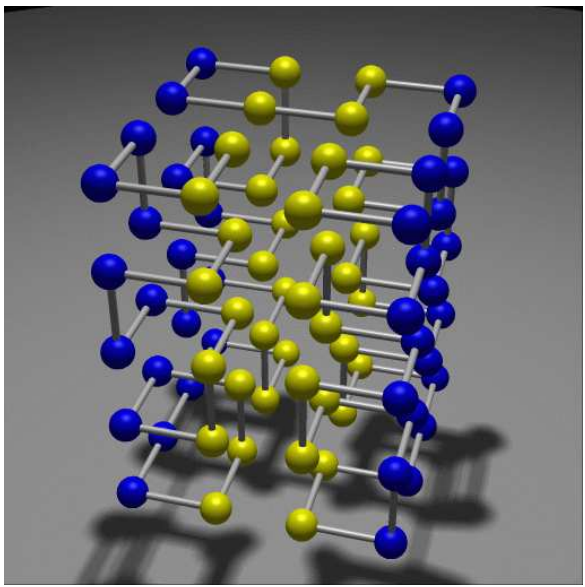, width=3truein}
  \end{center}
\caption{ Putative native state of the ``four helix bundle" 
  sequence, see Table~1, as proposed by O'Toole {\it et al.}. It has
  $E=-94$, fits into a rectangular box, and consists of three homogeneous
  layers. Structurally, it can be interpreted as four helix bundles.}
\label{3d.struct.old}
\end{figure}

A most interesting case is a 2-species 80-mer with interactions $(-1,0,-1)$
studied first in Ref.\cite{otoole}. These particular interactions were
chosen instead of the HP choice $(-1,0,0)$ because it was hoped that this
would lead to compact configurations. Indeed, the sequence was specially
designed to form a ``four helix bundle" which fits perfectly into a
$4\times4\times5$ box, see Fig.~\ref{3d.struct.old}.  Its energy in this
putative native state is $E=-94$.  Although O'Toole {\it et al.}\cite{otoole}
used highly optimized codes, they were not able to recover this state by
MC. Instead, they reached only $E=-91$. Supposedly, a different state with
$E=-94$ was found in Ref.\cite{pekney}, but figure 10 of this paper, which
is claimed to show this configuration, has a much higher value of $E$.
Configurations with $E=-94$ but slightly different from that in
Ref.\cite{otoole} and with $E=-95$ were found in Ref.\cite{deutsch} by means of
an algorithm similar to that in Ref.\cite{pekney}. For each of these low energy
states the author needed about one week of CPU time on a Pentium.

\subsection{3d, infinitly many  monomer types}

Sequences with $N=27$ and with continuous interactions were studied by Klimov
{\it et al.}\cite{Klimov:Thirumalai:96}. Interaction strengths were sampled
from Gaussians with fixed non-zero mean and fixed variance. These
$N(N-1)/2$ numbers were first attributed randomly to the monomer pairs,
then they were randomly permuted, using a Metropolis accept/reject strategy
with a suitable cost function, to obtain good folders.  Such ``breeding"
strategies to obtain good folders were also developed and employed by other
authors for various
models,\cite{Shakhnovich:93-94,Ebeling:Nadler:95-97,deutsch-kurosky:96} and
seem necessary to eliminate sequences which fold too slowly and/or
unreliably. It is believed that also during biological evolution
optimization processes took place with similar effects, so that actual
proteins are better folders than random amino sequences.

\section{Results}
\label{sec:results}

Let us now discuss our results.
All CPU times quote below refer to SPARC Ultra machines with 167 MHz.

\subsection{2d HP model}
\label{sec:results:2dHP}

\begin{figure}[t]
  \begin{center}
    \epsfig{figure=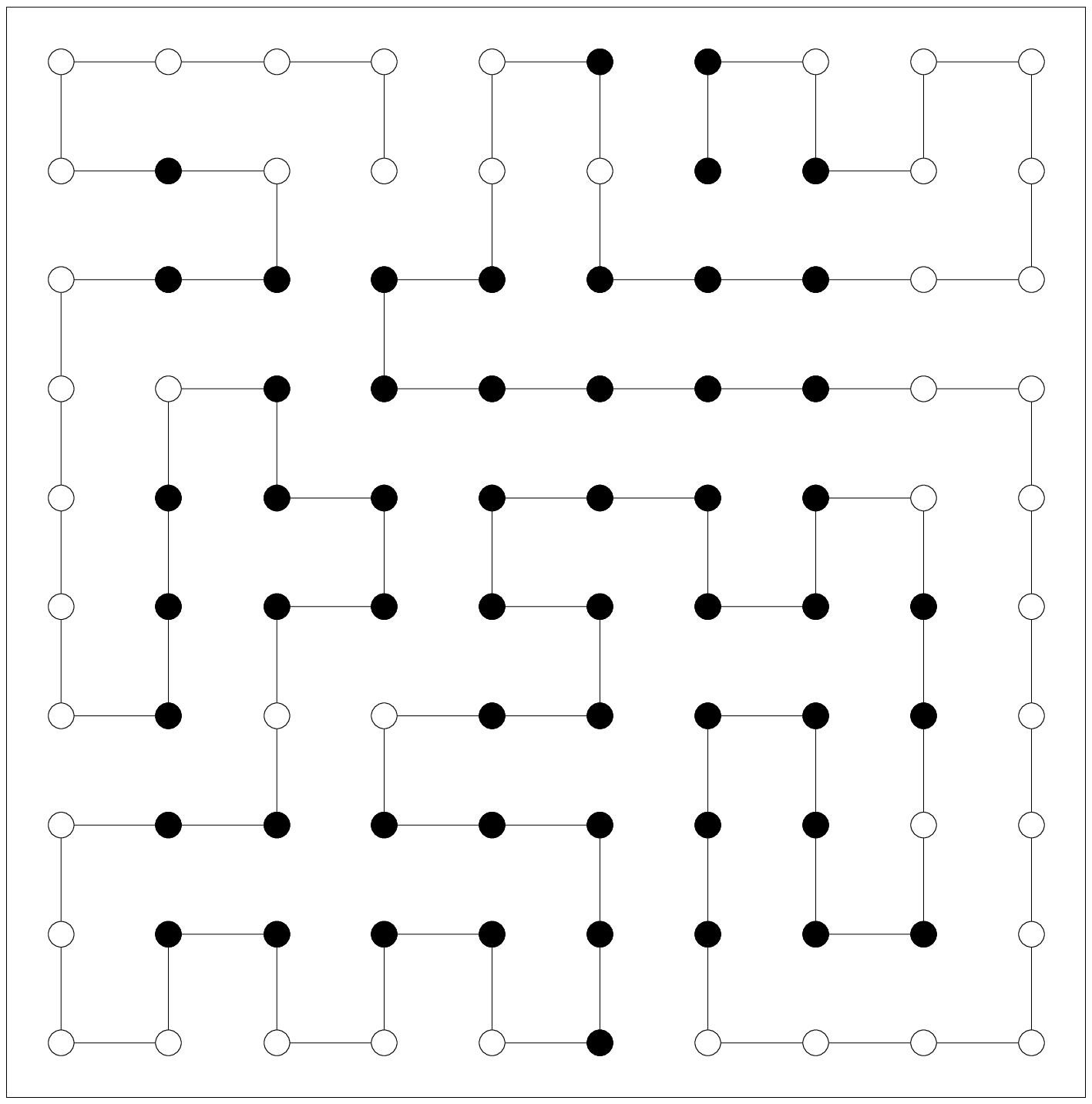, width=2.0truein}
    \hfil
    \epsfig{figure=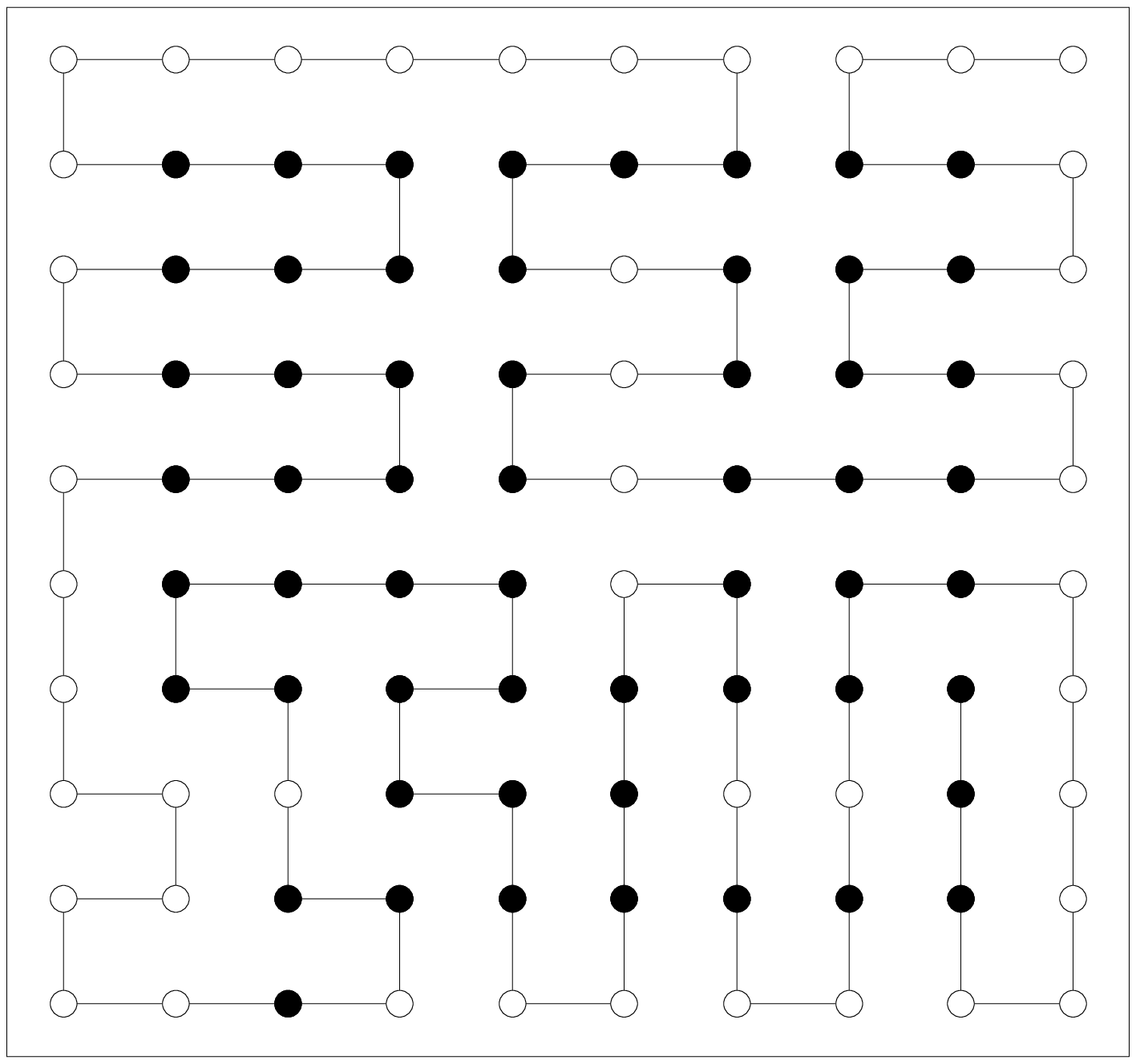, width=2.0truein}
  \end{center}
  \begin{center}
    \vglue-5mm
    \begin{minipage}[t]{6truecm}
      \caption{ One of the compact structures of sequence 1 with energy 
        ($E=-46$) lower than the ``native" state proposed by Ramakrishnan {\it et
          al.}\protect\cite{pekney}.\label{2d.seq1.struct.compact}}
    \end{minipage}
    \hfil
    \begin{minipage}[t]{6truecm}
      \caption{ One of the compact structures of sequence 2 with lower energy 
        ($E=-47$).\label{2d.seq2.struct.compact}}
    \end{minipage}
  \end{center}
  \begin{center}
    \vglue2mm
    \epsfig{figure=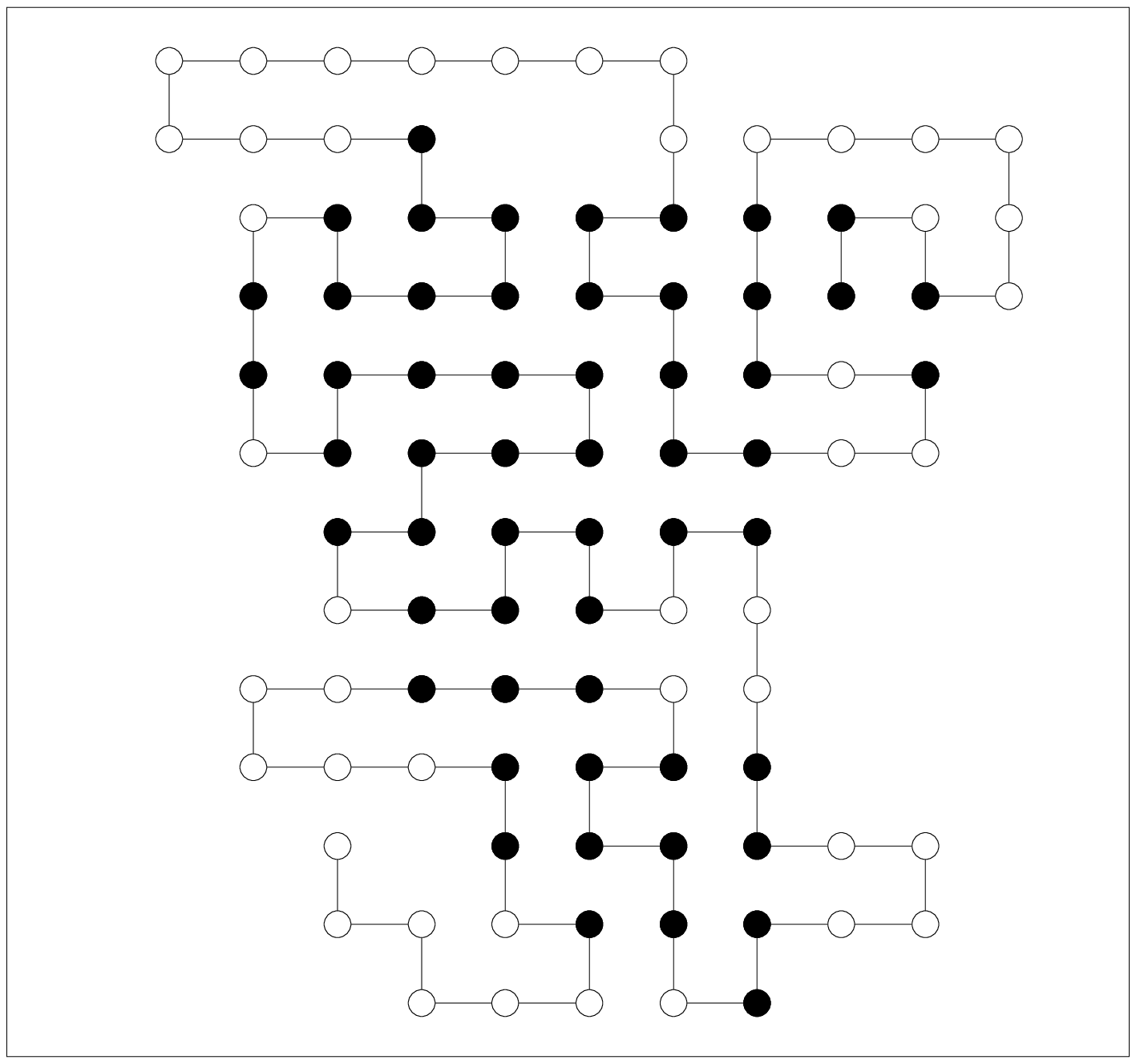, width=2.0truein}
    \hfil
    \epsfig{figure=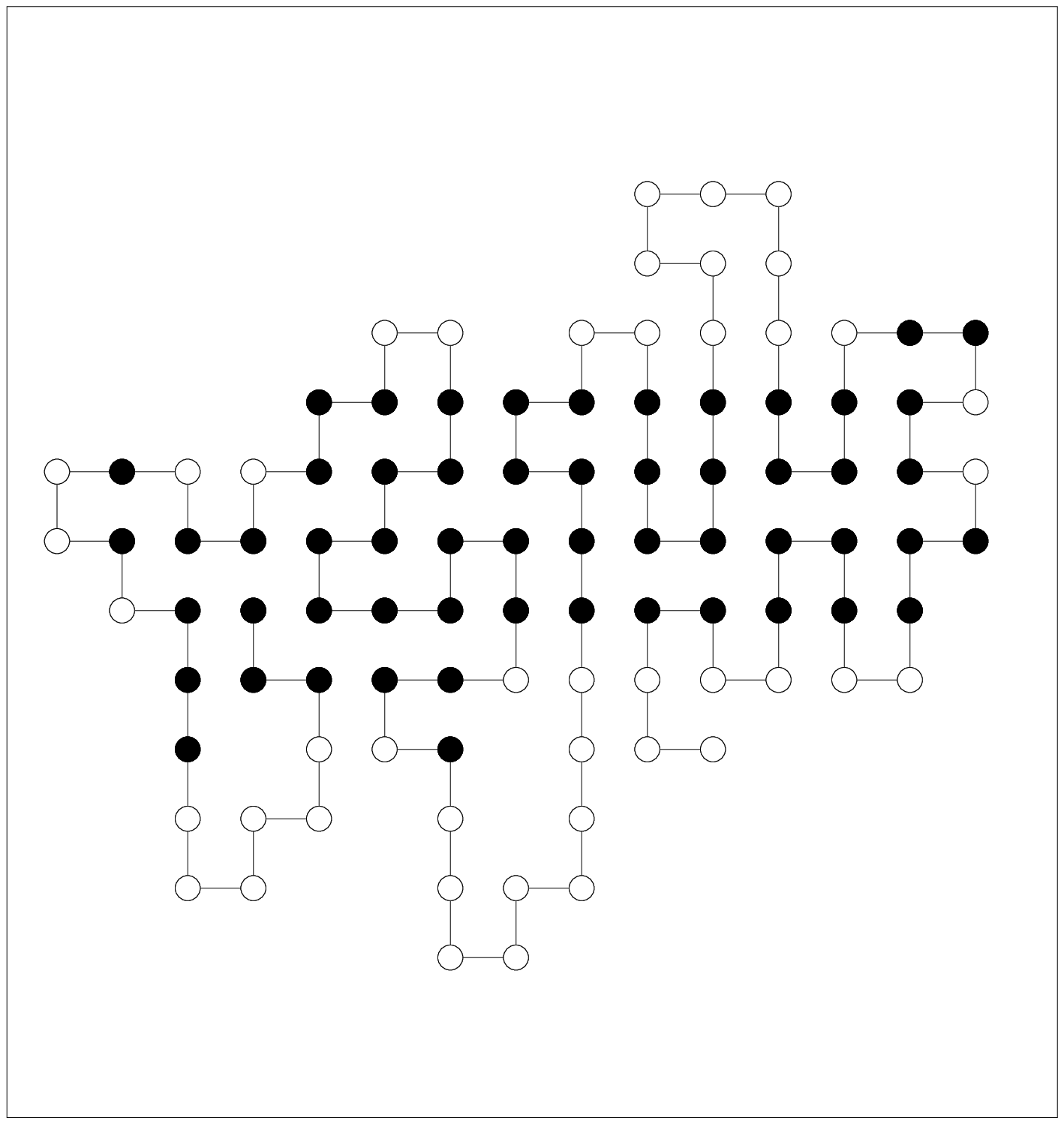, width=2.0truein}
  \end{center}
  \begin{center}
    \vglue-5mm
    \begin{minipage}[t]{6truecm}
      \caption{ One of the (non-compact) lowest energy sequences for sequence 1 
        ($E=-47$).\label{2d.seq1.struct.new}}
    \end{minipage}
    \hfil
    \begin{minipage}[t]{6truecm}
      \caption{ One of the (non-compact) lowest energy sequences for sequence 2 
        ($E=-49$).\label{2d.seq2.struct.new}}
    \end{minipage}
  \end{center}
\end{figure}

For the two HP chains of Ref.\cite{pekney} with $N=100$, see Table~1, we
found several compact states (within ca. 40 hours of CPU time) that had
energies lower than those of the compact putative ground states proposed in
Ref.\cite{pekney}. Figures \ref{2d.seq1.struct.compact} and
\ref{2d.seq2.struct.compact} show representative compact structures with
$E=-46$ for sequence 1 and $E=-47$ for sequence 2.  Moreover, we found
several non-compact configurations with energies even lower: the lowest
energies found within 1--2 days of CPU time had $E=-47$ and $E=-48$ for
sequence 1 and 2, respectively.  Forbidding non-bonded HP pairs, we
obtained even $E=-49$ for sequence 2. Figures \ref{2d.seq1.struct.new} and
\ref{2d.seq2.struct.new} show representative non-compact structures with
these energies; a non-exhaustive collection of these is listed in Table~1.
These results reflect the well-known property that HP sequences (and those
of other models) usually have ground states that are not maximally compact,
see, e.g. Yue {\it et al.}\cite{yue-shak}, although there is a persistent
prejudice to the contrary.\cite{otoole,pekney,Shakhnovich:Gutin:90-93}

\subsection{3d HP model}
\label{sec:results:3dHP}

\begin{figure}
\begin{center}
\epsfig{figure=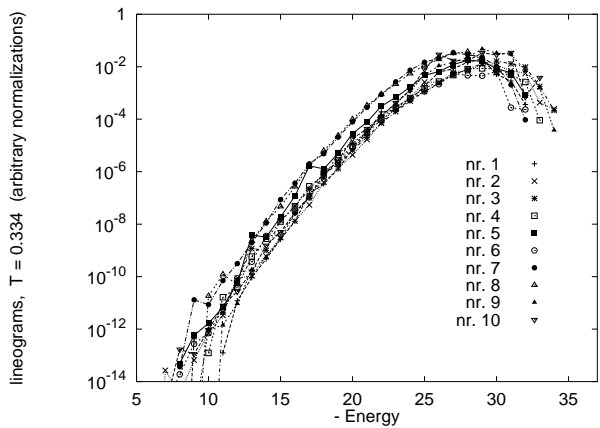, angle=270, width=8.3cm}
\end{center}
\vglue-3mm
\caption{Energy spectrum of the ten sequences given in Ref.\protect\cite{yue-shak}. 
More precisely, to emphasize the low-energy part of the spectrum, we show the 
histograms obtained from the spectra by multiplying 
with \protect$e^{E/T}, \;T=0.334$. Note that there are no 
energy gaps in any of these spectra.
}
\vglue-2mm
\label{NoGaps}
\end{figure}

With PERM we succeeded to reach ground states of the ten sequences of
length $N=48$ given in Ref.\cite{yue-shak} in {\it all} cases, in CPU times
between a few minutes and one day, see Table~2. In these simulations we
used a rather simple version of PERM, where we started assembly always from
the same end of the chain. We found that the sequences most difficult to
fold were also those which had resisted previous Monte Carlo
attempts.\cite{yue-shak} In those cases where a ground state was hit more
than once, we verified also that the ground states were highly degenerate.
In no case there were gaps between ground and first excited states, see
Fig.~\ref{NoGaps}.  Therefore, none of these sequences is a good folder,
though they were designed specifically for this purpose.

\begin{table}[t]
\caption{ PERM performance for the binary sequences from Ref.\protect\cite{yue-shak}
} \label{table2}
\begin{center}
\begin{tabular}{|ccccr|} \hline
 sequence & 
 ${-E_{\rm min}}^a$ &
 ${-E_{\rm MC}}^b$ & 
 ${n_{\rm success}}^c$ & CPU time   \\
   nr.    &                &                  &               &(min)  \\ \hline
    1     &    32    &    31    &   101     &     6.9  \\
    2     &    34    &    32    &    16     &    40.5  \\
    3     &    34    &    31    &     5     &   100.2  \\
    4     &    33    &    30    &     5     &   284.0  \\
    5     &    32    &    30    &    19     &    74.7  \\
    6     &    32    &    30    &    24     &    59.2  \\
    7     &    32    &    31    &    16     &   144.7  \\
    8     &    31    &    31    &    11     &    26.6  \\
    9     &    34    &    31    &     1     &  1420.0  \\
   10     &    33    &    33    &    16     &    18.3  \\\hline
\end{tabular}
\end{center}
\vglue1mm
\hglue2.1truecm$^a$ \footnotesize{Ground state energies.\protect\cite{yue-shak}}\\
\hglue2.1truecm$^b$ \footnotesize{Previously reached energies with Monte Carlo methods.\protect\cite{yue-shak}}\\
\hglue2.1truecm$^c$ \footnotesize{Number of independent tours in which a ground state was hit.}
\end{table}

\subsection{3d modified HP model}
\label{sec:results:3moddHP}

\begin{figure}[t]
  \begin{center}
    \epsfig{figure=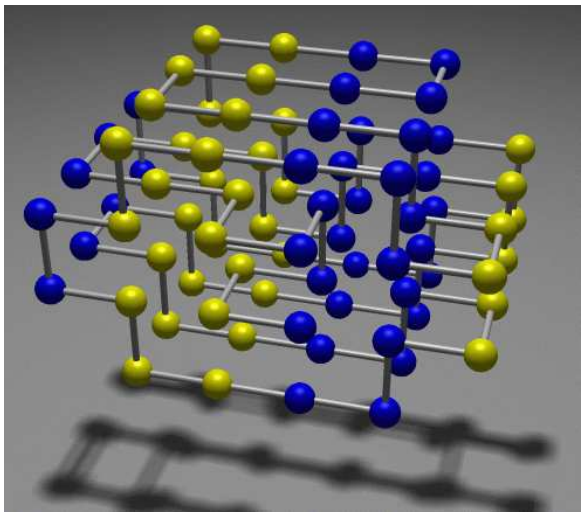, width=3truein}
  \end{center}
\caption{ Conformation of the ``four helix bundle"
sequence with $E=-98$. We propose that this is the actual ground 
state. Its shape is highly symmetric although it does not fit into a 
rectangular box. It is not degenerate except for a flipping of the central 
front $2\times2\times2$ box.}
\vglue-3mm
\label{3d.struct.new}
\end{figure}

For the two-species 80-mer with interactions $(-1,0,-1)$, even without much
tuning our algorithm gave $E=-94$ after a few hours, but it did not stop
there. After a number of rather disordered configurations with successively
lower energies, the final candidate for the native state has $E=-98$. It
again has a highly symmetric shape, although it does not fit into a
$4\times4\times5$ box, see Fig.~\ref{3d.struct.new}.  It has twofold
degeneracy (the central $2\times2\times2$ box in the front of
Fig.~\ref{3d.struct.new} can be flipped), and both configurations were
actually found in the simulations. Optimal parameters for the ground state
search in this model are $\beta=1/kT\approx 2.0$, $a_{PP} = a_{HH} \approx
2$, and $a_{HP}\approx -0.13$. With these, average times for finding
$E=-94$ and $E=-98$ in new tours are ca.~20 min and 80 hours, respectively.

A surprising result is that the monomers are arranged in four homogeneous 
layers in Fig.~\ref{3d.struct.new}, while they had formed only three 
layers in the putative ground state of Fig.~\ref{3d.struct.old}. 
Since the interaction should favor the segregation 
of different type monomers, one might have guessed that a configuration 
with a smaller number of layers should be favored. We see that this 
is outweighed by the fact that both monomer types can form large double 
layers in the new configuration. Again, our new ground state is 
not `compact' in the sense of minimizing the surface, and hence 
it also disagrees with the wide spread prejudice 
that native states are compact.

\begin{figure}
\begin{center}
\epsfig{figure=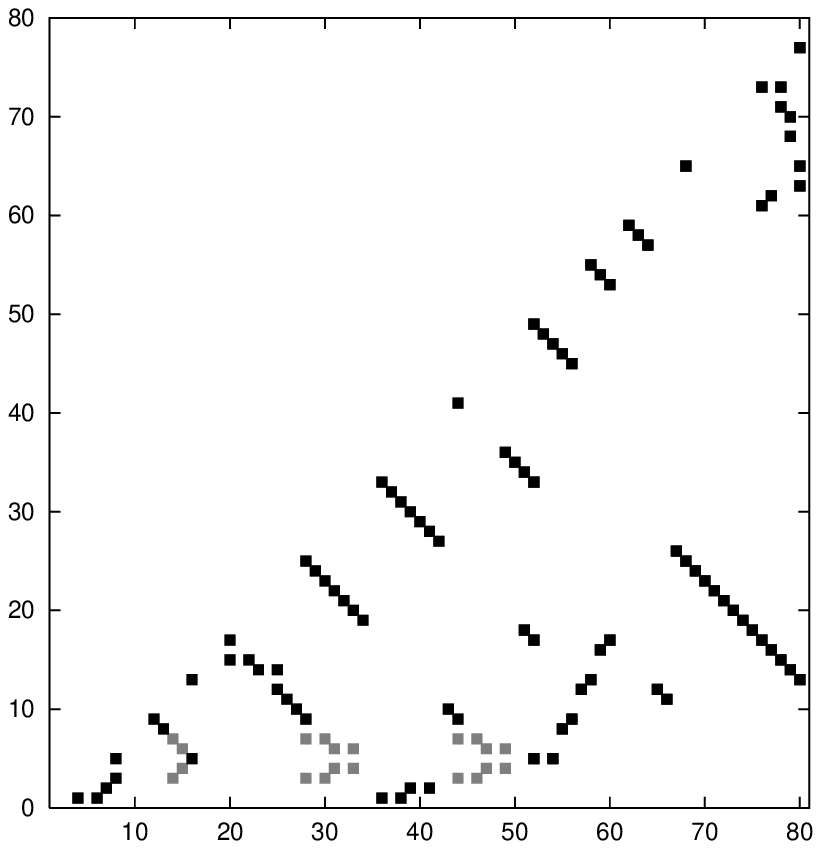, width=6truecm}
\hfil
\epsfig{figure=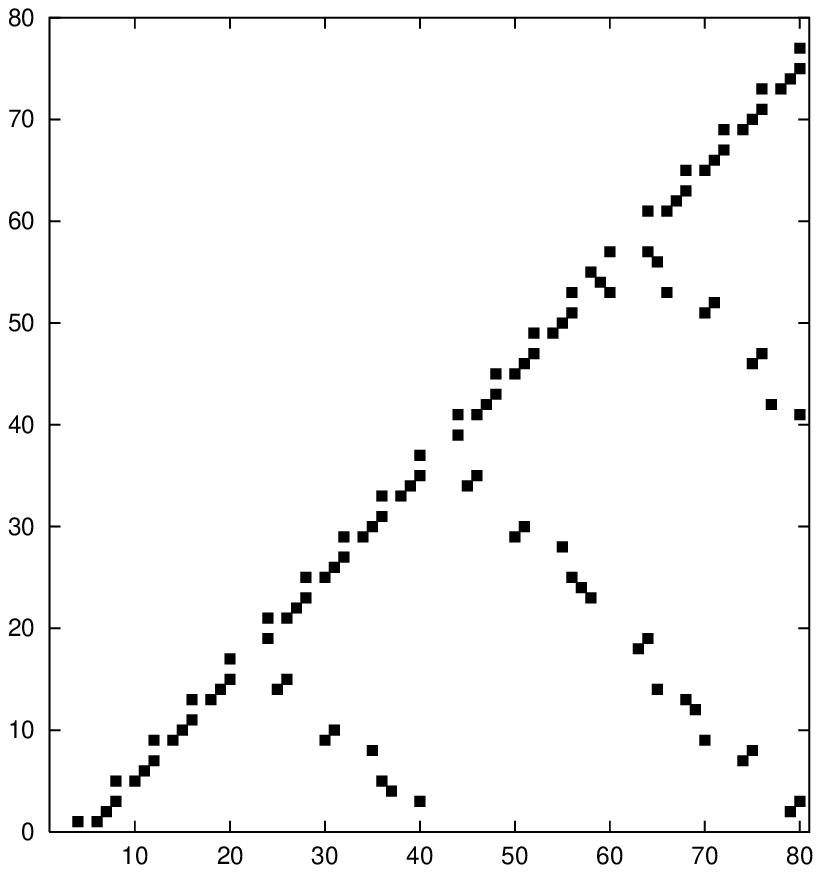, width=5.8truecm}
\end{center}
\begin{center}
  \vglue-5mm
  \begin{minipage}[t]{6.1cm}
    \caption{ Contact matrix of the structure in
      Fig.~\protect\ref{3d.struct.new}; a black point at $(i,j)$ indicates
      that there is a contact between monomer $i$ and monomer $j$; grey
      points indicate contacts in only one of the two native states,
      corresponding to the twofold degeneracy of the central
      $2\times2\times2$ box.  Note that the lines orthogonal to the main
      diagonal correspond to anti-parallel $\beta$-sheet secondary
      structure elements, see e.g. Ref.\protect\cite{Chan:Dill:89-90}.
      \label{3d.contact.new}}
  \end{minipage}
  \hfil
  \begin{minipage}[t]{5.7cm}
    \caption{ For comparison, the contact matrix of the putative ground
      state of Ref.\protect\cite{pekney} in
      Fig.~\protect\ref{3d.struct.old}; note that point triples close to
      the diagonal parallel as well as orthogonal to it are signatures of
      3d helical secondary structure elements, see e.g.
      Ref.\protect\cite{Chan:Dill:89-90}; the other points denote tertiary
      contacts between helices. \label{3d.contact.old}}
  \end{minipage}
\end{center}
\vglue2mm
  \begin{center}
    \epsfig{figure=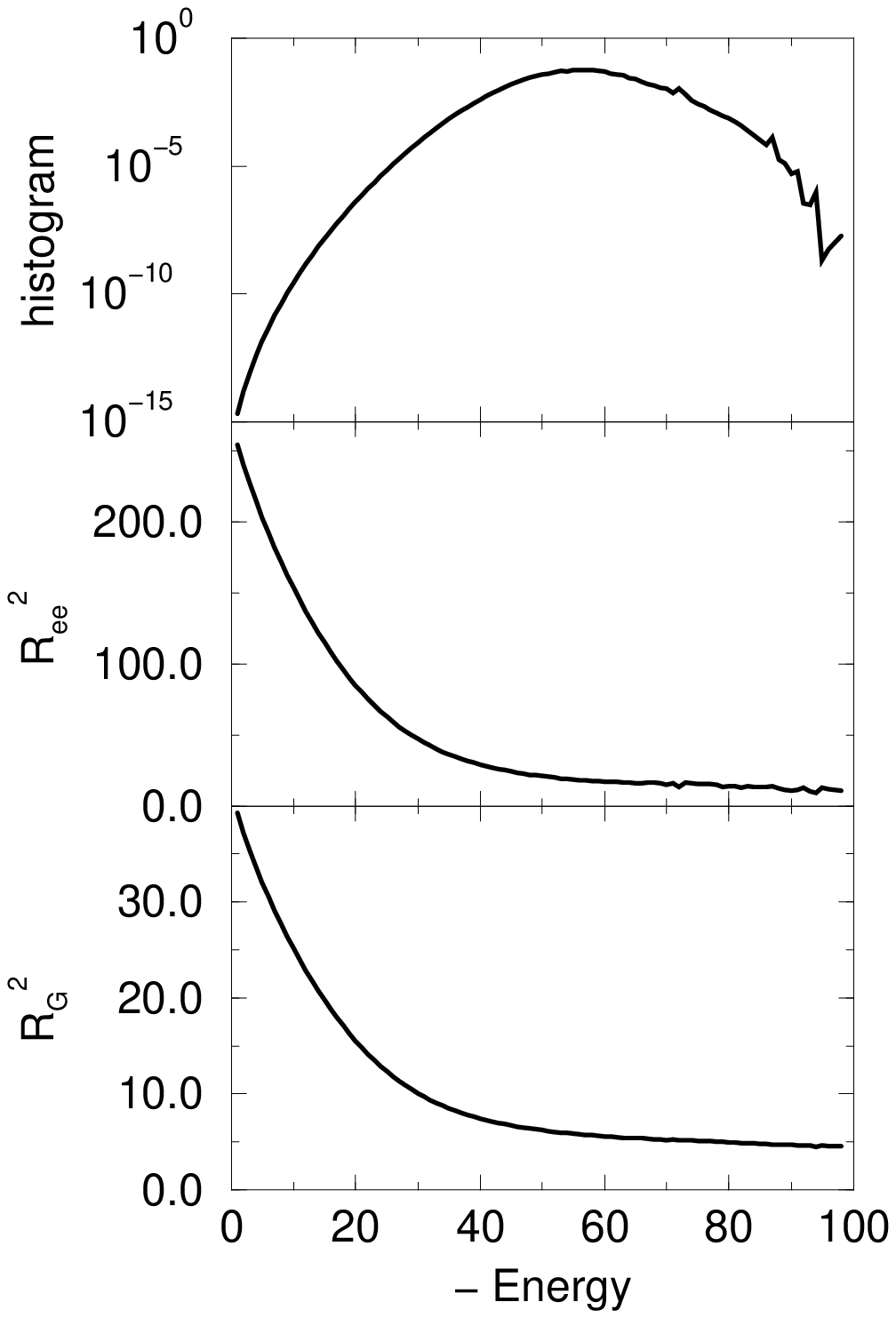, height=8truecm}
    \hfil
    \epsfig{figure=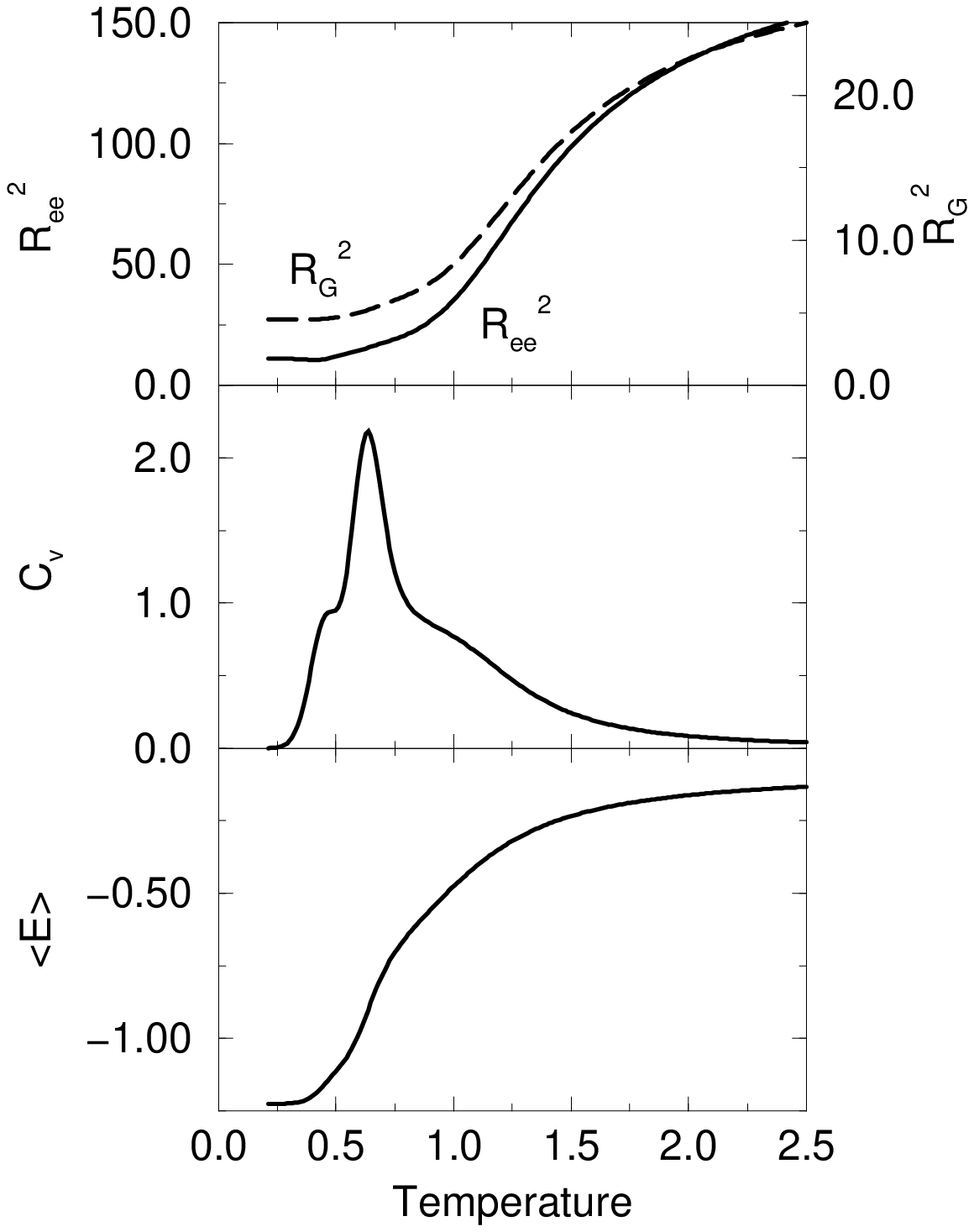, height=8truecm}
  \end{center}
  \begin{center}
  \vglue-8mm
    \begin{minipage}[t]{5.4truecm}
      \caption{Histograms of (top) thermal weight, (middle) radius of
        gyration, $R^2_G$, and (bottom) end-to-end distance, $R^2_{ee}$, $vs$
        energy $E$ for the 80-mer ``four helix bundle" at $T=0.75$.
        \label{hist.vs.e}}
    \end{minipage}
    \hfil
    \begin{minipage}[t]{6.6truecm}
      \caption{(top) Average end-to-end distance, $R^2_{ee}$, and radius of
        gyration, $R_G^2$, (middle) specific heat per monomer, $C_v$, and
        average energy per monomer, $<E>$, $vs$ temperature $T$ for the
        80-mer ``four helix bundle".\label{op.vs.T}}
    \end{minipage}
  \end{center}
  \vglue-5mm
\end{figure}

In terms of secondary structure, the new ground state is fundamentally
different from the putative ground state of Ref.\cite{pekney}. While the
new structure (Fig.~\ref{3d.struct.new}) is dominated by $\beta$ sheets,
which can most clearly be seen in the contact matrix (see
Fig.~\ref{3d.contact.new}), the structure in Fig.~\ref{3d.struct.old} is
dominated by helices, see also the corresponding contact matrix in
Fig.~\ref{3d.contact.old}.

\begin{figure}[t]
  \begin{center}
    \epsfig{figure=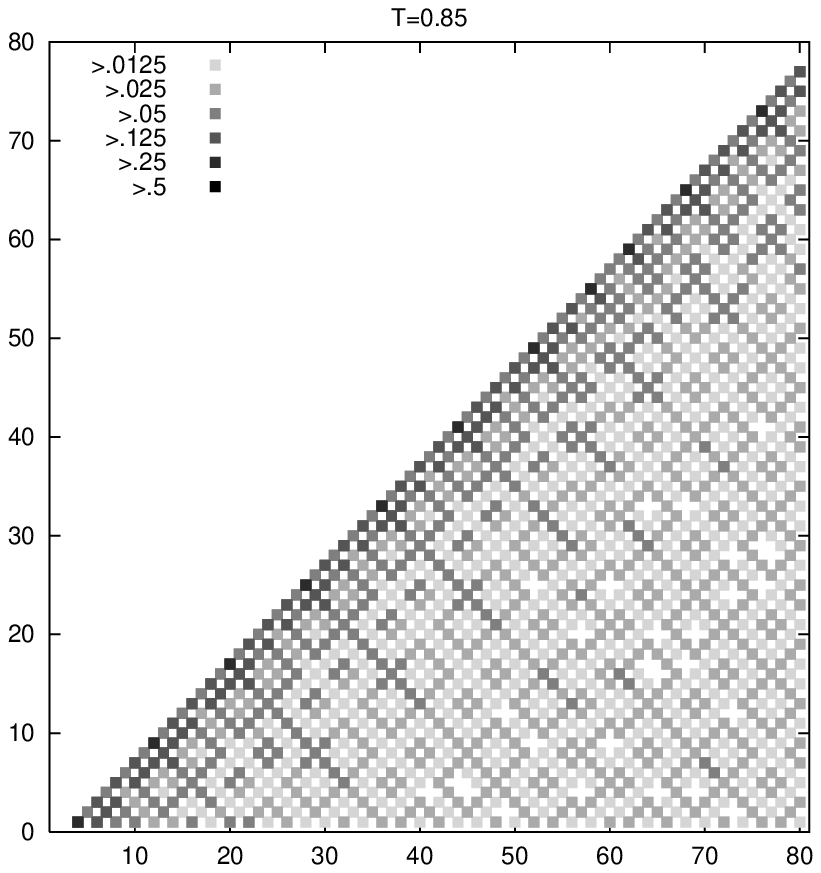, width=6truecm}
    \hfil
    \epsfig{figure=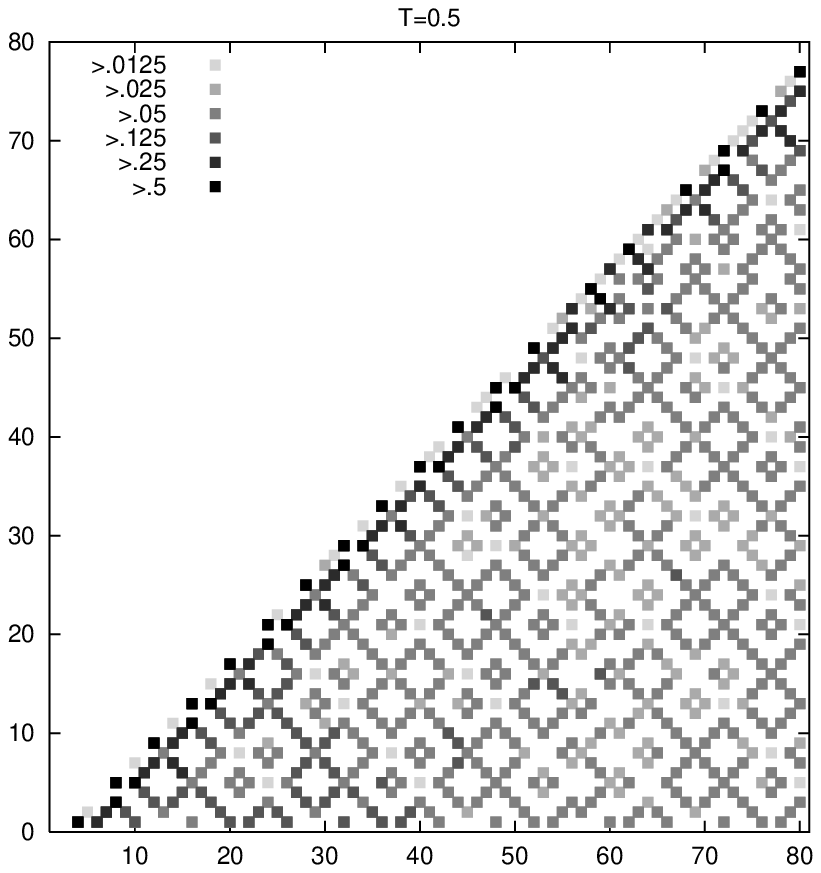, width=6truecm}
  \end{center}
  \begin{center}
    \begin{minipage}[c]{6truecm}
      \caption{Thermally averaged contact matrix for the 80-mer 
        ``four helix bundle" in the collapsed but unstructured phase
        ($T=0.85$).  \label{contact.85}}
    \end{minipage} 
    \hfil
    \begin{minipage}[c]{6truecm}
      \caption{Thermally averaged contact matrix for the 80-mer 
        ``four helix bundle" in the intermediate helix-dominated phase
        ($T=0.5$).\label{contact.50}}
    \end{minipage}
  \end{center}
  \begin{center}
    \epsfig{figure=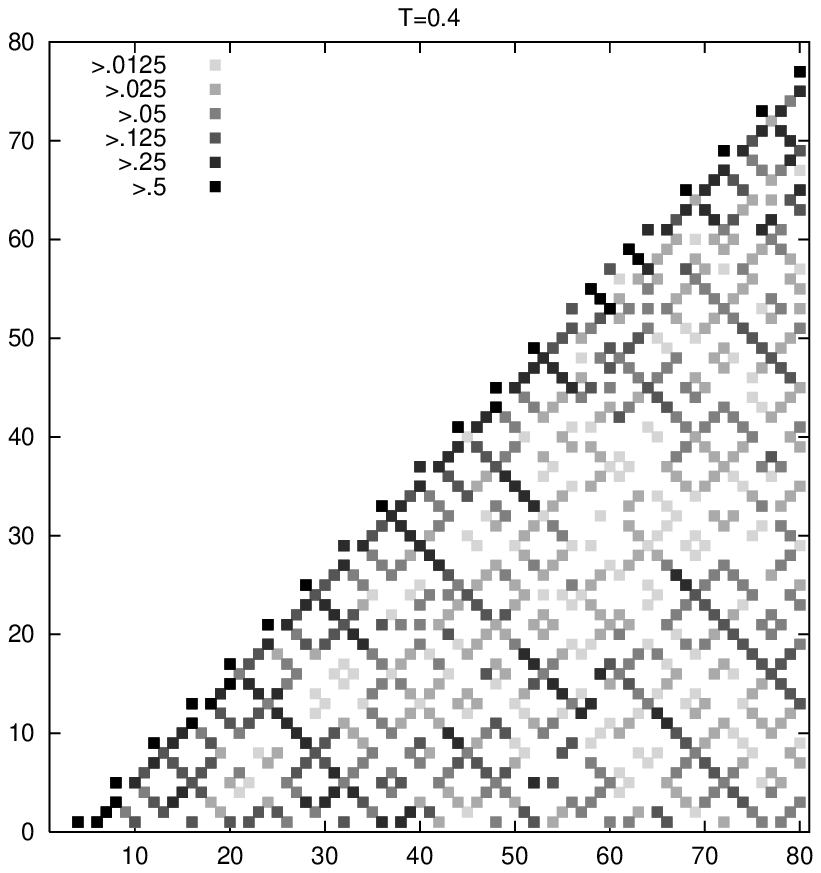, width=6truecm}
    \hfil
    \epsfig{figure=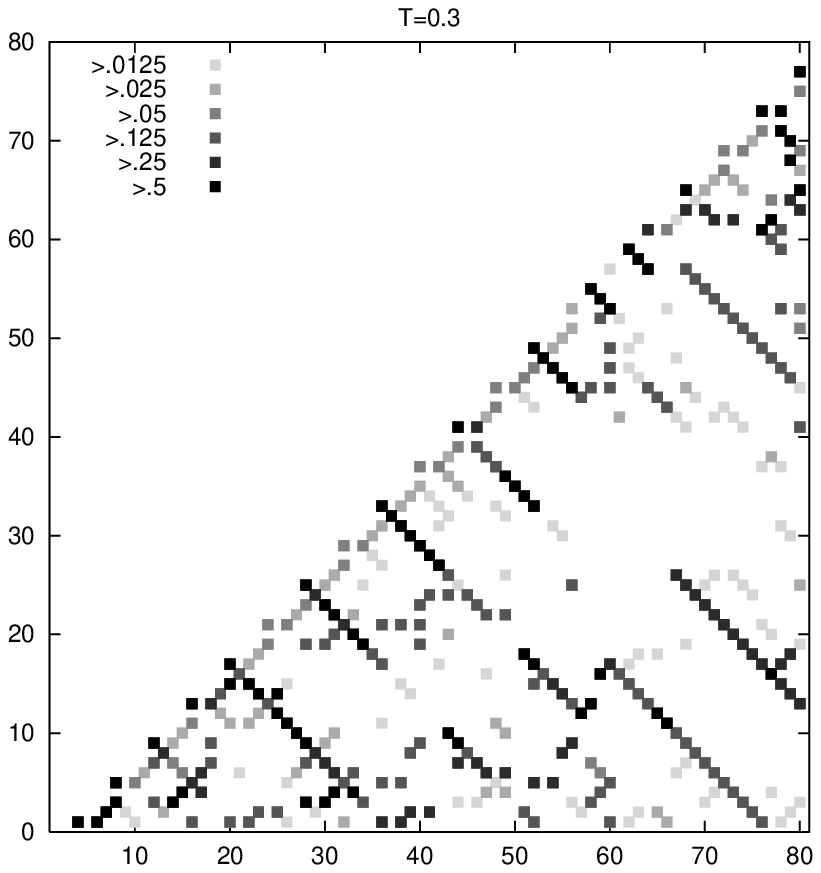, width=6truecm}
  \end{center}
  \begin{center}
    \vglue-5mm
    \begin{minipage}[t]{6truecm}
      \caption{Thermally averaged contact matrix for the 80-mer 
        ``four helix bundle" at the transition from the intermediate
        helix-dominated to the $\beta$-sheet dominated phase
        ($T=0.4$). \label{contact.40}}
    \end{minipage} 
    \hfil
    \begin{minipage}[t]{6truecm}
      \caption{Thermally averaged contact matrix for the 80-mer 
        ``four helix bundle" in the $\beta$-sheet dominated phase
        ($T=0.3$).\label{contact.30}}
    \end{minipage}
  \end{center}
  \vglue-5mm
\end{figure}
In order to analyze the folding transition of this sequence we again
constructed histograms of the distribution of energy, end to end distance,
and radius of gyration, by combining the results obtained at various
temperatures between $T=0.45$ and 3.  Figure~\ref{hist.vs.e} shows these
distributions, reweighted so that it corresponds to $T=0.75$.  The thermal
behavior of these order parameters as functions of $T$ is obtained by
Laplace transform, and is shown in Fig.~\ref{op.vs.T}.  The behavior of
energy, end-to-end distance and radius of gyration follow closely each
other and exhibit only the smeared out collapse of the chain from a
random coil to some unstructured compact phase. In contrast, the specific
heat exhibits more structure: the shoulder around $T=1$ corresponds again
to the coil-globule collapse, but there are additional transitions seen
around $T=0.62$ and $T=0.45$. The last one is the transition to the
$\beta$-sheet dominated native phase. However, the transition at $T=0.62$
is from a unstructured globule to an intermediate phase that is
helix-dominated but exhibits strong tertiary fluctuations. These
structural transitions are illustrated in Figs~\ref{contact.85} to
\ref{contact.30} where the thermally averaged contact matrices are shown
for the respective phases.

The intermediate, helix-dominated phase is particularly interesting.
To it apply some of the usual characteristics of a molten globule 
state\cite{Pain:93}: i) compactness, ii) large secondary structure content 
(although not necessarily native), and iii) strong fluctuations.

\subsection{3d, infinitly many  monomer types}
\label{sec:results:3dinfinity}

For all sequences with $N=27$ from Ref.\cite{Klimov:Thirumalai:96} we could
reach the supposed ground state energies within $< 1$ hour.  In no case we
found energies lower than those quoted in Ref.\cite{Klimov:Thirumalai:96},
and we verified also the energies of low-lying excited states given in
Ref.\cite{Klimov:Thirumalai:96}.  Notice that these sequences were designed
to be good folders by Klimov {\it et al.}\cite{Klimov:Thirumalai:96}.  This
time the design had obviously been successful, which is mainly due to the
fact that the number of different monomer types is large.  All sequences
showed some gaps between the ground state and the bulk of low-lying states,
although these gaps are not very pronounced in some cases.

More conspicuous than these gaps was another feature: all low lying 
excited states were very similar to the ground state, as measured by
the fraction of contacts which existed also in the native configuration.
Stated differently, if the gaps were not immediately obvious, this was 
because they were filled by configurations which were very similar to 
the ground state and can therefore easily transform into the native state 
and back. Such states therefore cannot prevent a sequence from being 
a good folder. For none of the sequences of Ref.\cite{Klimov:Thirumalai:96} 
we found truely misfolded low-lying states with small overlap with the 
ground state.

\begin{figure}[b]
  \begin{center}
    \epsfig{figure=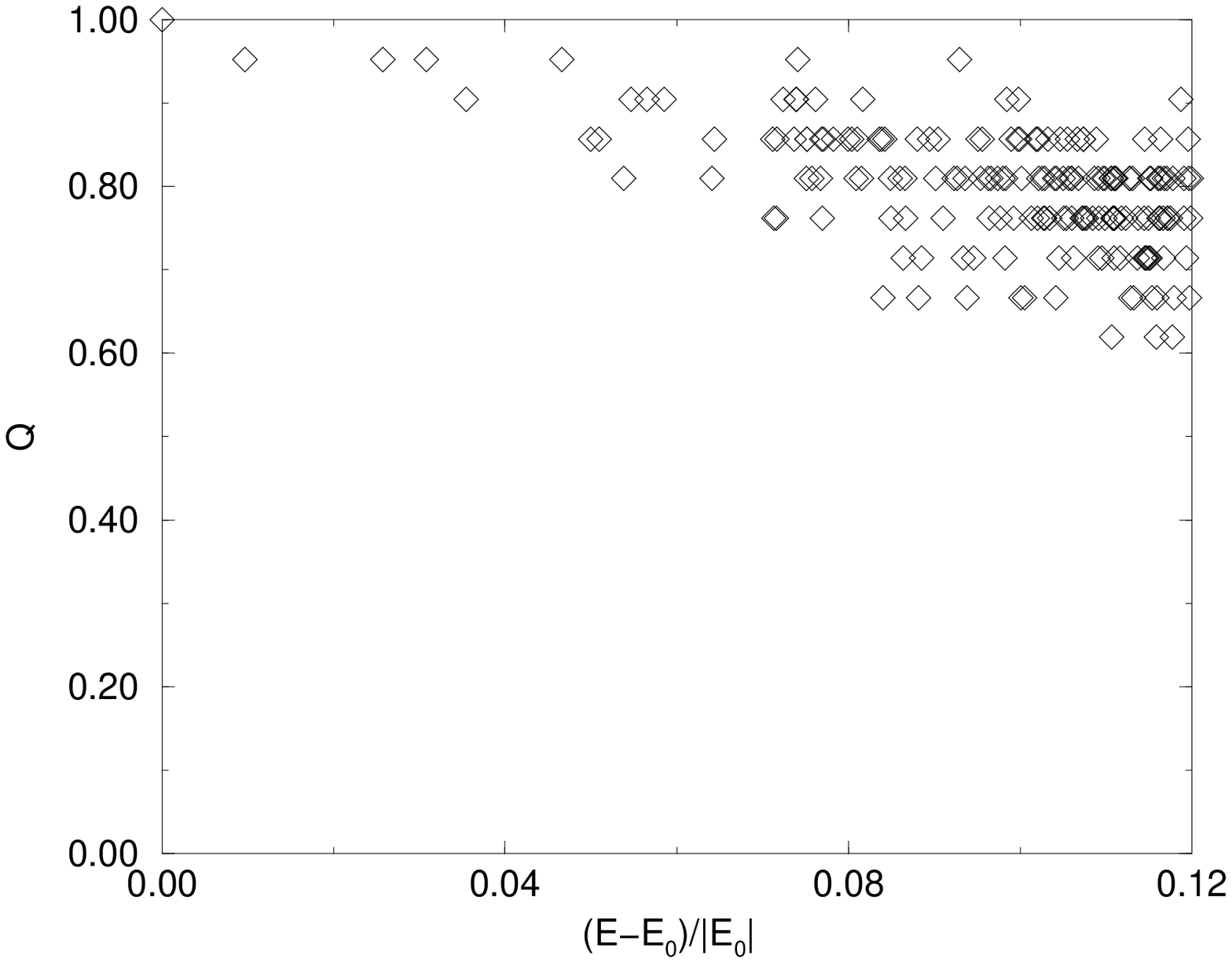, width=6truecm}
    \hfil
    \epsfig{figure=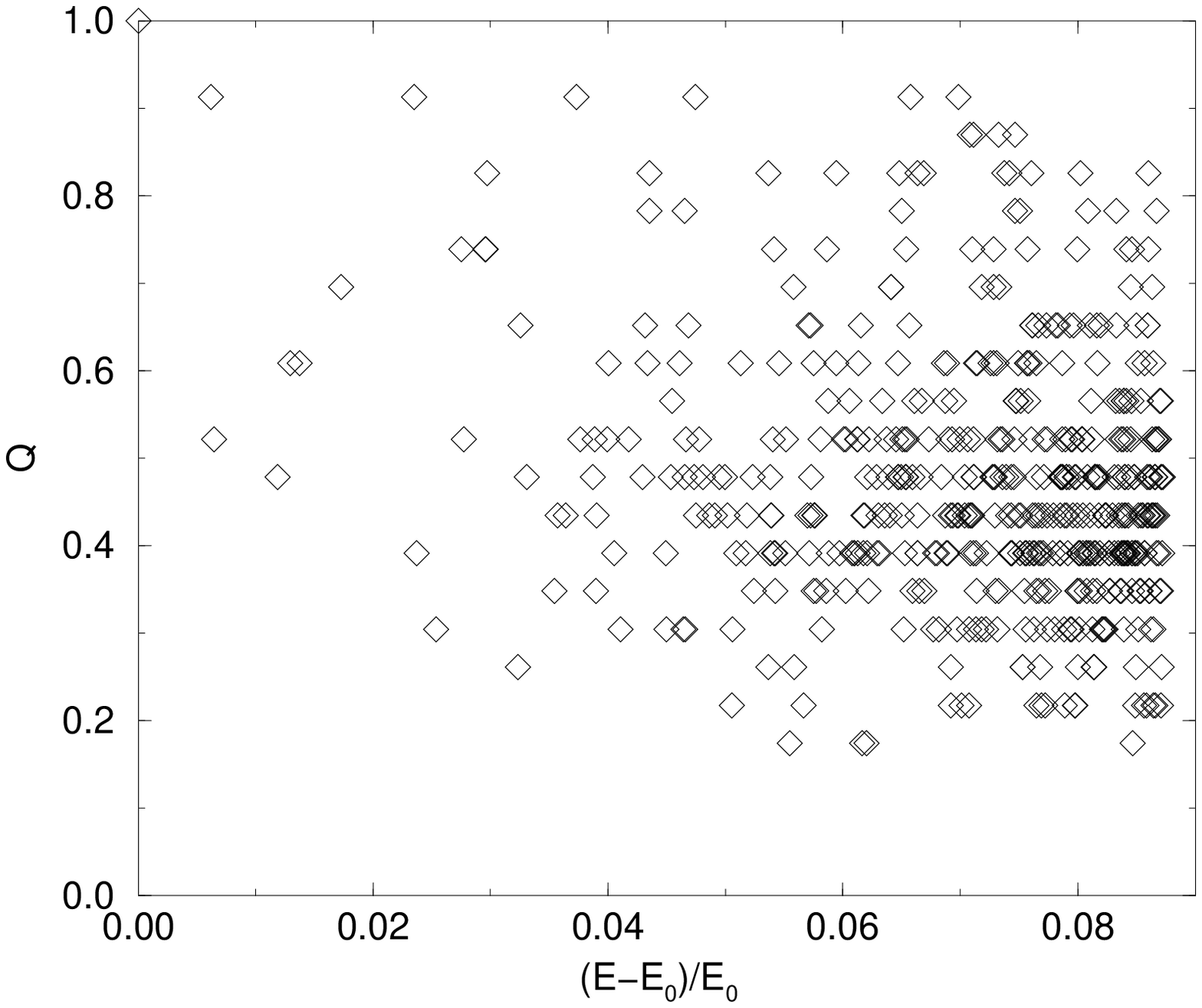, width=5.58truecm}
  \end{center}
  \begin{center}
    \vglue-7mm
    \begin{minipage}[t]{5.9truecm}
      \caption{ Overlap with ground state, $Q$, $vs$ energy, $E$, of the
        lowest energy conformations for sequence no. 70 of
        Ref.\protect\cite{Klimov:Thirumalai:96}. \label{klimov.Q.vs.E.single} }
    \end{minipage}
    \hfil
    \begin{minipage}[t]{5.8truecm}
      \caption{ Overlap with ground state, $Q$, $vs$ energy, $E$, of the
        lowest energy conformations for a single random sequence.
        \label{klimov.Q.vs.E.random.single} }
    \end{minipage}
  \end{center}
\end{figure}

Figure~\ref{klimov.Q.vs.E.single} illustrates this feature for one particular 
sequence. There we show the {\it overlap} $Q$, defined as the fraction 
by non-bonded nearest-neighbor ground state contacts which exist also in 
the excited 
state, against the excitation energy. For this and for each of the following 
figures, the 500 lowest-lying states were determined. We see no low energy 
state with a small value of $Q$. To demonstrate that this is due to 
design, and is not a property of random sequences with the same potential 
distribution, we show in Fig.~\ref{klimov.Q.vs.E.random.single} the analogous 
distribution for a random sequence. 

\begin{figure}
  \begin{center}
    \epsfig{figure=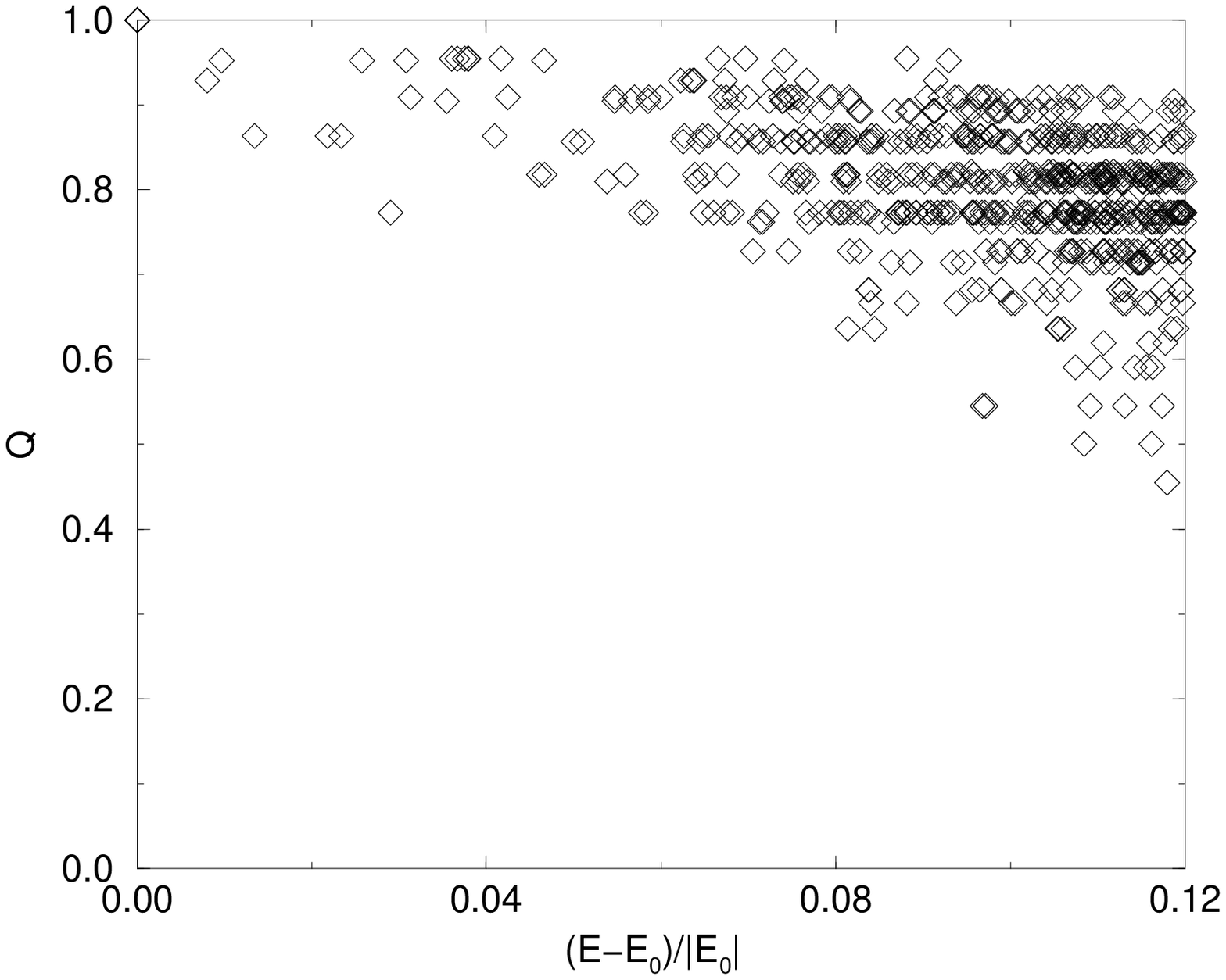, width=6truecm}
    \hfil 
    \epsfig{figure=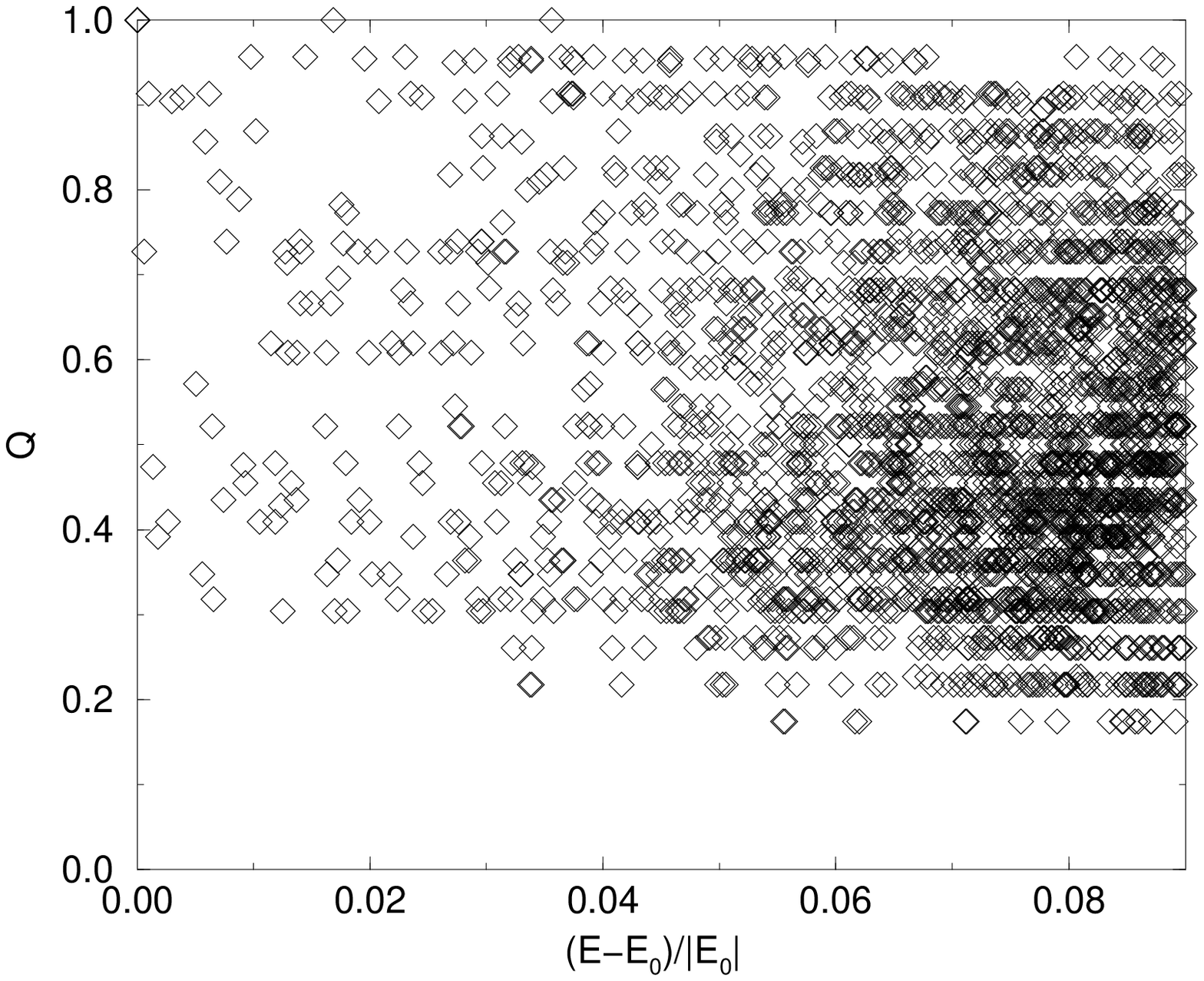,width=5.8truecm}
  \end{center}
  \begin{center}
    \vglue-7mm
    \begin{minipage}[t]{5.9truecm}
      \caption{ Overlap with ground state, $Q$, $vs$ energy, $E$, of the
        lowest energy conformations for sequences no. 61 to 70 of
        Ref.\protect\cite{Klimov:Thirumalai:96}; for better visibility, the
        same symbol is used for all sequences. \label{klimov.Q.vs.E.total}}
    \end{minipage}
    \hfil
    \begin{minipage}[t]{5.8truecm}
      \caption{ Overlap with ground state, $Q$, $vs$ energy, $E$, of the
        lowest energy conformations for ten random sequences; for better
        visibility, the same symbol is used for all sequences.
        \label{klimov.Q.vs.E.random.total}}
    \end{minipage}
  \end{center}
\end{figure}

To demonstrate that this difference is not merely due to a statistical
fluctuation, we show in Fig.~\ref{klimov.Q.vs.E.total} the distributions
for ten sequences from Ref.\cite{Klimov:Thirumalai:96} collected in a
single plot. Since the ground state energies differ considerably for
different sequences, we used normalized excitation energies $(E-E_0)/E_0$
on the x-axis. Analogous results for ten random sequences are shown in
Fig.~\ref{klimov.Q.vs.E.random.total}.  While there is no obvious
correlation between $Q$ and excitation energies for the random case, all
low energy states with small $Q$ have been eliminated in the designed
sequences.

This elimination of truely misfolded low energy states without elimination
of native-like low energy states might be an unphysical property of the
design procedure used in Ref.\cite{Klimov:Thirumalai:96}, but we do not
believe that this is the case. Rather, it should be a general feature of
any design procedure, including the one due to biological evolution. It
contradicts the claim of Ref.\cite{Sali:Shakhnovich:Karplus:94:Nature} that
it is only the gap between native and first excited state which determines
foldicity. On the other hand, our results are consistent with the ``funnel"
scenario for the protein folding
process,\cite{Wolynes:Onuchic:Thirumalai:95} where the folding pathway
consists of states successively lower in energy and closer to the native
state.

We note that for random sequences there are also excited 
states that have unit overlap with the native state, a feature not 
present in the folding sequences. These are cases where the native state 
has open loops and/or dangling ends, so that more compact conformations
have all contacts of the native state, but have---in addition---energetically 
unfavorable contacts resulting in a higher total energy.

\section{Summary and Outlook}
\label{sec:summary}

We showed that the pruned-enriched Rosenbluth method (PERM) can be very
effectively applied to protein structure prediction in simple lattice
models. It is suited for calculating statistical properties and is very
successful in finding native states. In all cases it did better than any
previous MC method, and in several cases it found lower energy states than
those which had previously been conjectured to be native.

We verified that ground states of the HP model are highly degenerate and
have no gap, leading to bad folders.  For sequences that are good folders
we have established a funnel structure in state space: low-lying excited
states of well-folding sequences have strong similarities to the ground
state, while this is not true for non-folders with otherwise similar
properties.

Especially, we have presented a new candidate for the native configuration
of a ``four helix bundle" sequence which had been studied before by several
authors.  The ground state structure of the ``four helix bundle" sequence,
being actually $\beta$-sheet dominated, differs strongly from the
helix-dominated intermediate phase.  This sequence, therefore, should not
be a good folder.

Although we have considered only lattice models in this paper, we should
stress that this is not an inherent limitation of PERM.  Straightforward
extensions to off-lattice systems are possible and are efficient for
homopolymers at relatively high temperatures.\cite{alg} Recently performed
simulations of a 2-dimensional off-lattice protein model
\cite{irbaeck2d:97} have shown that PERM is not quite as fast as the {\it
  simulated tempering\/} \cite{parisi:92} algorithm used by Irb\"ack {\it
  et al.}\cite{irbaeck2d:97}. But there are several ways to improve PERM.
One way could be to adopt a strategy similar to simulated tempering.
Another improvement of PERM which could be particularly useful for
off-lattice simulations might consist in more sophisticated algorithms for
positioning the monomers when assembling the chain. In particular one might
try to use some feedback of successfully build up low energy states or
subchains for the positioning of previous monomers in the following tours.
Finally, one can use configurations reached by PERM as a starting point for
alternative (Metropolis or greedy) searches.  Work along these lines is in
progress, and we hope to report on it soon.

\vspace*{-2pt}
\section*{Acknowledgments}

The authors are grateful to Gerard Barkema for helpful discussions 
during this work. One of them (P.G.) wants to thank also Eytan Domany 
and Michele Vendruscolo for very informative discussions, and to 
Drs. D.K. Klimov and R. Ramakrishnan for correspondence.


\section*{References}
\begin{thebibliography}{99}
  
\bibitem{Gierasch:King:90} L. M. Gierasch and J. King (eds.), 
{\it Protein Folding, Deciphering the Second Half of the Genetic Code} 
(AAAS, New York, 1990)

\bibitem{Creighton:92} T.E. Creighton (ed.), {\it Protein Folding} (Freeman, 
  New York, 1992)

\bibitem{Merz:LeGrand:94} 
K. M. Merz Jr. and S. M. LeGrand (eds.), 
{\it The Protein Folding Problem and Tertiary Structure Prediction} 
(Birkh\"auser, Boston, 1994)

\bibitem{Brunak:95} 
H. Bohr and S. Brunak (eds.), {\it Protein Folds: A Distance Based Approach}
(CRC Press, Boca-Raton/FL. 1996).

\bibitem{Drexler:86} 
{K.~E.~Drexler},~{\it Engines~of~Creation}, (Anchor Books, 1986); 
available also on the web at the URL
{\sf http://www.asiapac.com/EnginesOfCreation/\/}.

\bibitem{Gross:95} 
M. Gro\ss, {\it Expeditionen in den Nanokosmos}, (Birkh\"auser, Basel, 1995).

\bibitem{Crippen:Maiorov:94} 
G. M. Crippen and V. N. Maiorov, in Ref.\protect\cite{Merz:LeGrand:94}, p.231--277.

\bibitem{Kolinski:Skolnick:95} 
A. Kolinski and J. Skolnick, {\it Lattice Models of Protein Folding,
Dynamics and Thermodynamics},
(Chapman \& Hall, New York, 1996).

\bibitem{Mirny:Shakhnovich:96}
L. A. Mirny and E. I. Shakhnovich, J. Mol. Biol. {\bf 264}, 1164 (1996).

\bibitem{Sali:Shakhnovich:Karlplus:94:JMB}
A. Sali, E. I. Shakhnovich and M. Karplus J. Mol. Biol. {\bf 235}, 1614 (1994).

\bibitem{Hansmann:Okamoto:94-96}
U. H. E. Hansmann and Y. Okamoto, J. Comp. Chem. {\bf 14}, 1333 (1993);
Physica A {\bf 212}, 415 (1994); Phys. Rev. E {\bf 54}, 5863 (1996).

\bibitem{Wilson:Cui:94}
S. R. Wilson and W. Cui, in Ref.\protect\cite{Merz:LeGrand:94}, p.43--70.

\bibitem{Unger:Moult:93} R. Unger and J. Moult, J. Mol. Biol. {\bf 231}, 75 
(1993)

\bibitem{LeGrand:Merz:94a}
S. M. LeGrand and K. M. Merz jr., in Ref.\protect\cite{Merz:LeGrand:94}, p.109--124.

\bibitem{Dill:Fiebig:Chan:93}
K. A. Dill, K. M. Fiebig and H. S. Chan, Proc. Natl. Acad. Sci USA {\bf 90}, 
1942 (1993).

\bibitem{Eisenhaber:Persson:Argos:95}
F. Eisenhaber, B. Persson and P. Argos,
Crit. Rev. Biochem. Mol. Biol. {\bf 30}, 1 (1995).

\bibitem{alg} P.~Grassberger, Phys. Rev. {\bf E 56}, 3682 (1997).
  
\bibitem{rr} M.N.~Rosenbluth and A.W.~Rosenbluth, J.~Chem.~Phys. {\bf 23}, 256 
(1955)
  
\bibitem{kantor-kardar:94} Y. Kantor and M. Kardar, Europhys. Lett. 
   {\bf 28}, 169 (1994)  

\bibitem{Dill:85} K.A.~Dill, Biochemistry {\bf 24}, 1501 (1985) 

\bibitem{Dill:89-91} K.F.~Lau and K.A.~Dill, Macromolecules {\bf 22}, 
  3986 (1989); J. Chem. Phys. {\bf 95}, 3775 (1991); 
  H.S.~Chan, and K.A.~Dill, J. Chem. Phys. {\bf 95}, 3775 (1991)

\bibitem{socci1} N.D.~Socci and J.N. Onuchic, J.~Chem.~Phys. {\bf 101},
  1519 (1994)

\bibitem{otoole} E.~O'Toole and A.~Panagiotopoulos, J.~Chem.~Phys.
  {\bf 97}, 8644 (1992)

\bibitem{Sali:Shakhnovich:Karplus:94:Nature}
A. Sali, E. I. Shakhnovich and M. Karplus, Nature {\bf 369} 248 (1994).

\bibitem{Klimov:Thirumalai:96} D.K. Klimov and D. Thirumalai, Proteins:
  Structure, Function and Genetics {\bf 26}, 411 (1996); sequences
  available from {\sf http://www.glue.umd.edu/$\sim$klimov}

\bibitem{kremer} J. Batoulis and K. Kremer, J. Phys. {\bf A 21}, 127 (1988) 

\bibitem{garel} T. Garel and H. Orland, J. Phys. {\bf A 23}, L621 (1990)

\bibitem{velicson} B. Velikson, T. Garel, J.-C. Niel, H. Orland and 
     J.C. Smith, J. Comput. Chem. {\bf 13}, 1216 (1992)

\bibitem{umrigar} C.J. Umrigar, M.P. Nightingale, and K.J. Runge, 
  J. Chem. Phys. {\bf 99}, 2865 (1993) 

\bibitem{enrich} F.T. Wall and J.J. Erpenbeck, J. Chem. Phys. {\bf 30},
  634, 637 (1959)

\bibitem{multic} H.~Frauenkron and P.~Grassberger,
  J.~Chem.~Phys.~\textbf{107}, 9599 (1997)

\bibitem{stiff} U.~Bastolla and P.~Grassberger, J.~Stat.~Phys.~\textbf{89},
   1061 (1997)

\bibitem{Matheson:Scheraga:78}
R. R. Matheson and H. A. Scheraga, Macromolecules {\bf 11}, 819 (1978).

\bibitem{Grassberger:unpublished} At first sight, one might believe that
  allowing the chain to grow at both ends should decrease the attrition
  rate and hence be advantageous. For homopolymers one can see easily that
  this is not true. Attrition is actually decreased, since chains which
  have one end blocked can still grow at the other end. But such chains
  have small Rosenbluth factors and thus extremely low $W$ in average.
  Therefore, they just cost efforts without efficiently contributing to
  statistical averages.  For heteropolymers this argument no longer holds,
  however, and it is not clear why growing the chain at both ends is not
  efficient either.

\bibitem{Finkelstein:Gutin:Badretdinov:95}
A. V. Finkelstein, A. M. Gutin and A. Y. Badretdinov
PROTEINS: Structure, Function and Genetics {\bf 23}, 151 (1995).

\bibitem{Frauenkron:etal:97} 
H. Frauenkron, U. Bastolla, E. Gerstner, P. Grassberger 
and W. Nadler,  Phys. Rev. Lett. {\bf 80}, 3149 (1998).

\bibitem{proteins:97} 
U. Bastolla, H. Frauenkron, E. Gerstner, P. Grassberger 
and W. Nadler, to appear in PROTEINS (1998).

\bibitem{pekney} R.~Ramakrishnan, B.~Ramachandran and J.F.~Pekney,
  J.~Chem.~Phys. {\bf 106}, 2418 (1997)

\bibitem{yue-shak} K.~Yue {\it et al.}, Proc.~Natl.~Acad.~Sci.~USA
  {\bf 92}, 325 (1995)

\bibitem{Shakhnovich:93-94}
E. I. Shakhnovich and A. M. Gutin, Proc. Natl. Acad. Sci USA {\bf 90}, 7195 
(1993);
E. I. Shakhnovich, Phys. Rev. Lett. {\bf 72}, 3907 (1994).

\bibitem{Yue:Dill:95}
K. Yue and K. A. Dill, Proc. Natl. Acad. Sci USA {\bf 92}, 146 (1995).

\bibitem{deutsch} J.M.~Deutsch, J. Chem. Phys. {\bf106}, 8849 (1997).
  
\bibitem{Ebeling:Nadler:95-97} M. Ebeling and W. Nadler,
Proc. Natl. Acad. Sci. USA {\bf 92}, 8798 (1995);
Biopolymers {\bf 41}, 165 (1997).

\bibitem{deutsch-kurosky:96} J.M. Deutsch and T. Kurosky, 
Phys. Rev. Lett. {\bf 76}, 323 (1996).

\bibitem{Shakhnovich:Gutin:90-93}
E. I. Shakhnovich and A. M. Gutin, J. Chem. Phys. {\bf 93}, 5967 (1990);
A. M. Gutin and E. I. Shakhnovich, J. Chem. Phys. {\bf 98}, 8174 (1993).

\bibitem{Chan:Dill:89-90}
H. S. Chan and K. A. Dill,
Macromolecules {\bf 22}, 4559 (1989);
J. Chem. Phys. {\bf 92}, 3118 (1990).

\bibitem{Pain:93}
H. Christensen and R. H.Pain,
Europ. Biophys. J. {\bf 19}, 221 (1991).

\bibitem{Wolynes:Onuchic:Thirumalai:95}
P. G. Wolynes, J. N. Onuchic, and D. Thirumalai,
Science {\bf 267}, 1619 1995).

\bibitem{irbaeck2d:97} A.~Irbaeck, C.~Peterson and F.~Potthast,
  J.~Phys.~Rev.~{\bf E 55}, 860 (1997).

\bibitem{parisi:92} E.~Marinari and G.~Parisi, Europhys.~Lett.~{\bf 19}, 451
  (1992).
\end{thebibliography}
\end{document}